\documentclass[prb,twocolumn]{revtex4}

\usepackage{amsmath, amsthm, amssymb}

\newcommand{\rss}{\rm\scriptscriptstyle}

\newcommand{\ds}{\displaystyle}

\usepackage{graphicx}
\usepackage{psfrag}

\begin{document}

\title{Surface Melting of the Vortex Lattice in Layered
Superconductors: \\Density Functional Theory}
\author{A.\ De\ Col$^{1}$, G.I.\ Menon$^{2}$, 
V.B.\  Geshkenbein$^{1,3}$, and G.\ Blatter$^{1}$}
\affiliation{$^{1}$Theoretische Physik, ETH-Z\"urich, 
CH-8093 Z\"urich, Switzerland}
\affiliation{$^{2}$The Institute of Mathematical Sciences, 
C.I.T. Campus, Taramani, Chennai 600\ 113, India }
\affiliation{$^3$L.D.\ Landau Institute for Theoretical 
Physics RAS,   117940 Moscow, Russia}
\date{\today}


\begin{abstract}

We study the effects of an $ab$-surface on the vortex-solid 
to vortex-liquid transition in layered superconductors
in the limit of vanishing inter-layer Josephson coupling. 
We derive the interaction between pancake vortices in a 
semi-infinite sample and adapt the density functional theory
of freezing to this system. We obtain an effective one-component 
order-parameter theory which can be used to describe the 
effects of the surface on vortex-lattice melting. 
Due to the absence of protecting layers in the neighbourhood of
the surface, the vortex lattice formed near the surface
is more susceptible to thermal fluctuations. 
Depending on the value of the magnetic field, we predict
either a continuous or a 
discontinuous surface melting transition.
For intermediate values of the magnetic field, 
the surface melts continuously, assisting 
the formation of the liquid phase and suppressing hysteresis
above the melting transition, a prediction consistent
with experimental results.  For very low and very high magnetic
fields, the surface melts discontinuously.
The two different surface melting 
scenarios are separated by two surface multicritical points, which
we locate on the melting line. 

\end{abstract}

\maketitle


\section{Introduction}

The melting of the vortex lattice is one
of the most remarkable aspects of the phenomenology of
the high-temperature superconductors\cite{nelson88,houghton89,brandt89,sengupta91,zeldov95,schilling96}.
Such a melting transition, in common with other discontinuous phase
transitions, should involve metastable phases such
as an undercooled liquid and an overheated solid.
Experiments on the mixed phase in the layered 
high-T$_c$ superconductor BSCCO however indicate
an asymmetric hysteresis at the melting transition
\cite{soibel00}, involving a supercooled 
liquid phase but no overheated solid. 
In a recent letter\cite{decol05b}, we studied the
impact of an $ab$-surface on the vortex lattice melting 
in layered superconductor and we found that the surface 
can have a profound effect on the hysteretic behaviour.
Our approach combined ideas of 
density functional theory with a mean-field substrate model proposed in 
Ref.\ \onlinecite{dodgson00}. The present paper provides 
detailed derivations of the results outlined in 
Ref.\ \onlinecite{decol05b}. In addition, we present 
several new results, such as a detailed analysis of 
the solid-liquid interface and of the multi-critical point 
at low magnetic fields.

\begin{figure}[h]
\centering
   \includegraphics [width= 8cm] {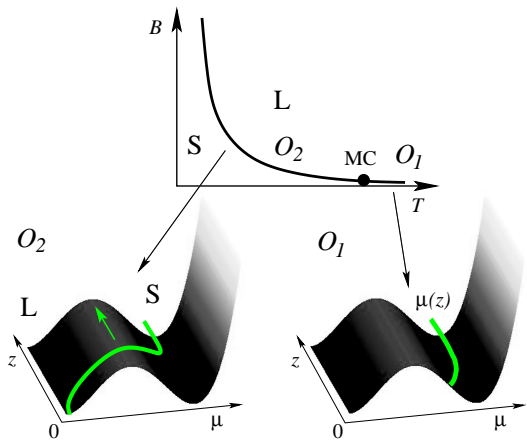}
   \caption{Top: schematic phase diagram of the pancake
   vortex system. The multicritical point (MC) separates the 
   portion of the melting line where the surface melts continuously, 
   $O_2$ transition (left), from the region where the surface 
   melts discontinuously, $O_1$ transition (right). 
   Lower left: configuration of the 
   order-parameter profile $\mu(z)$ for an $O_2$ transition at the 
   melting temperature $T_\mathrm{m}$. 
   In the language of Landau theory we use in the paper, the 
   problem maps to the determination of the minimal energy configuration
   of an elastic string, $\mu(z)$, subject to a three-dimensional 
   potential landscape, which is described by the bulk free energy, 
   plus an additional surface term, which pushes 
   the tip of the string towards the liquid minimum. 
   In an $O_2$ transition the surface 
   promotes the formation of the liquid phase upon 
   approaching $T_\mathrm{m}$ from below.
   The order-parameter profile reaches the liquid minimum 
   continuously at the surface for $T\nearrow T_\mathrm{m}$, while 
   it remains in the `solid' well in the bulk. 
   The liquid propagates continuously from
   the surface into the bulk, thus removing the solid phase.
   Lower right: configuration of the order-parameter profile in 
   an $O_1$ transition at melting. 
   The order parameter reaches a finite value on 
   the surface at $T_\mathrm{m}$. Melting occurs via a finite 
   jump for all $z$ and the solid phase can be overheated. 
}
\label{fig:main}
\end{figure}

The absence of a metastable solid phase in ordinary
solids has been the object of great interest in the literature
\cite{Ubbelohde1978,pietronero79,Lipowsky1982,Frenken1985,
Trayanov1987,Tartaglino2005}.  It is now widely
accepted that such asymmetric hysteresis can be
related to the action of surface (pre-)melting.
Surfaces act as a natural nucleation point for
the liquid phase and inhibit the appearance of the
solid at temperatures above melting.  
This is due to the fact that the atoms at the surface 
experience a weaker stabilizing potential.  
Indeed, the stability analysis for a semi-infinite atom chain
\cite{pietronero79} shows that the 
enhanced motion of the first atoms makes 
the `surface' unstable at a lower
temperature with respect to the bulk. This suggests
that the surface may represent a favorable nucleation
site for the liquid phase. However, criteria based on the stability of
the solid phase cannot  address the state of the
system beyond the surface instability point.

A comprehensive description of the melting transition in the 
presence of a surface requires the inclusion of the effects of both the 
solid and the liquid phases on an
equal footing\cite{Trayanov1987,Trayanov1988,Ohnesorge1994}.
On a phenomenological level, this can be done 
within the framework of Landau theory using
the order parameters of the melting transition, 
i.e., the Fourier components of the particle density.
The destabilizing effect of the surface can be accounted
for\cite{Lipowsky1982,lipowsky83} by a surface 
term favoring the appearance of the liquid phase.
Following the discussion in Ref.\ \onlinecite{lipowsky83}, 
as the simplest example of such a theory 
one considers a Landau free energy of the type
\begin{equation}\label{landau_s}
 F[\mu] =\! \int_0^{\infty} \!\!\! dz\, 
 \Bigl[\frac{\ell^2}{2}\Bigl(\frac{d\mu}{dz}\Bigr)^2 
 + f(\mu) + \ell \delta(z)f_1(\mu)\Bigr],
\end{equation}
where $z$ is the distance from the surface, $\mu$ the 
($z$-dependent) order parameter and $f(\mu)$ and 
$f_1(\mu)$ the bulk free energy and surface destabilizing 
potential respectively. For a bulk first-order phase transition
with an unstable surface we can choose a form
\begin{align}
 f(\mu)     & = (a/2)\mu^2 -(b/3)\mu^3+(c/4)\mu^4, \nonumber \\ 
 f_1(\mu) & = (a_1/2)\mu^2, \nonumber
\end{align}
where the coefficients $a_1, b, c$ are chosen to be strictly positive 
and temperature independent, whilst temperature enters in 
$a(T)$ via the usual Ansatz for a first-order phase transition 
$a(T) - a_\mathrm{m} \propto T - T_\mathrm{m}$, 
with $T_\mathrm{m}$ the melting temperature and 
$a_\mathrm{m} = 2b^2/9c$.

The bulk free energy $f$ shows two minima: one at $\mu_\mathrm{l} = 0$ 
(liquid) with energy $f_\mathrm{l}=0$ and a second one 
at $\mu_\mathrm{s}>0$ (solid), whose 
energy depends on temperature. The liquid is the stable minimum 
above the bulk transition temperature $T_\mathrm{m}$, while below 
$T_\mathrm{m}$ the solid becomes the stable phase. At the transition
the two minima assume the same energy 
and the two phases can coexist. 

In the absence of a surface term, 
the stable configuration is given by a constant order parameter profile
throughout the sample. However, the surface free 
energy $f_1$ breaks the invariance of the system along $z$ and 
pushes the tip of the order parameter profile towards the liquid minimum 
at $\mu_\mathrm{l}=0$. This term competes with
the elastic energy term $\propto (d\mu/dz)^2$. 
A quantitative understanding of the surface destabilization requires to solve
the saddle-point equation for the free energy (\ref{landau_s}). 
The Euler-Lagrange equation provides the differential equation
\begin{equation}\label{eq_land}
 \ell^2 \frac{d^2\mu(z)}{dz^2} = \frac{\partial f(\mu(z))}{\partial \mu(z)}
\end{equation}
which has to be supplemented with the boundary condition originating 
from the surface term,
\begin{equation}\label{bc_land}
  \ell \frac{d \mu(z)}{dz}\Bigr|_{z=0} = 
   \frac{\partial f_1(\mu(z))}{\partial \mu}\Bigr|_{\mu(z= 0)} 
  =a_1 \mu(z=0).
\end{equation}
Two qualitatively different scenarios are possible
depending on the strength of the surface term or, more 
precisely\cite{lipowsky83}, on the ratio $a_1^2/a_\mathrm{m}$.
For relatively large perturbations ($a_1^2/a_\mathrm{m} \gg 1$) 
the surface assists the formation of the liquid at the melting transition. 
Approaching $T_\mathrm{m}$ from below, the 
bulk order parameter remains in the `solid' well 
(see Fig.\ \ref{fig:main}), while at the surface the order parameter
approaches the liquid minimum at the transition 
temperature {\it continuously}. The liquid phase then propagates 
from the surface and the overheated solid is removed above 
$T_\mathrm{m}$. Thus, even if the order parameter 
in the bulk jumps discontinuously, the surface melts continuously;
as the transition is second order at the surface, 
in Ref.\ \onlinecite{lipowsky83} it is referred to as an $O_2$ transition.
On the other hand, when the surface free energy is weak 
($a_1^2/a_\mathrm{m} \ll 1$), the order parameter at the surface 
jumps {\it discontinuously} to the liquid minimum across the
transition as in the bulk, although from a reduced value (see Fig.\ \ref{fig:main}). 
In Ref.\ \onlinecite{lipowsky83}, this discontinuous surface melting is 
called an $O_1$ transition since the transition is first order at the surface.  
In this case, the surface does not inhibit the appearance of an 
overheated solid phase.

In this paper, we determine the effects of the surface on the melting 
of the vortex system, starting from a `microscopic' theory which 
accounts for the modification of the inter-vortex interactions at the 
surface. Our program is to derive an appropriate Landau-type 
free energy similar to Eq.\ (\ref{landau_s}) by considering 
a coarse grained theory where the modulation of the thermally averaged 
vortex density plays the role of the order parameter. The Density Functional 
Theory (DFT) of freezing\cite{ramakrishnan79} provides an appropriate 
tool for such a study as it is based on the change of the free energy from 
a uniform liquid to a modulated state 
\cite{sengupta91,menon96,cornaglia00,De-Col2006,Curtin1989,Ohnesorge1994}.  
The only input which is needed in the analysis is the direct correlator, which 
we derive within  the substrate model approach\cite{dodgson00}. Restricting
ourselves to periodic modulations, one obtains a free-energy functional in 
terms of the Fourier component of the pancake vortex density at the first 
reciprocal lattice vectors of the ordered lattice. Since these are the natural 
order parameters of the melting transition, the DFT approach provides 
a microscopic derivation of a Landau type description of the transition 
similar to Eq.\ (\ref{landau_s}).

As a result of our DFT analysis, we find that for very low and very high 
magnetic fields the surface order-parameter still undergoes a finite,
although reduced, jump at the transition ($O_1$).  On the other hand,
for intermediate values of $B$, stray magnetic fields destabilize the 
layers close to the surface, leading to a continuous transition at the 
surface ($O_2$). Consequently, for a wide range of experimentally 
accessible fields we expect that surface fluctuations are sufficiently 
strong to assist the smooth propagation of the liquid phase into the 
bulk and to prevent the appearance of the metastable overheated 
solid phase;  this result is consistent with the experiments of Soibel 
{\em et al.}\cite{soibel00}. The surface continuous and discontinuous 
transitions are separated by a multi-critical point. While a 
precise location of the high-field multi-critical point goes beyond the 
limit of validity of our analysis, we have located the low-field multi-critical 
point by means of an analytical solution of the DFT equations. This 
analysis is further confirmed numerically.

The paper is organized as follows: we begin by calculating the modified 
in-plane and out-of-plane pancake vortex interactions (Section II) for a
semi-infinite system. In Section III, we review the classic DFT of freezing
and derive a specific functional which describes the vortex system in a 
layered superconductor.  We briefly describe the results of the novel DFT-substrate 
approach for an infinite (bulk) system (Section IV), where other results are 
available for comparison. We turn to the problem with the surface in Section V, 
where we present an analytical solution of the problem and show that depending 
on the value of the magnetic field the system exhibits either a 
`surface non-melting' ($O_1$) or a `surface melting' ($O_2$) behavior. 
Finally, in Section \ref{sec:surfnum} we confirm the validity of our analytical 
approach by a direct numerical solution of the DFT equations.
 

\section{Model}\label{sec:model}

\begin{figure}[t]
\centering
   \includegraphics [width= 7.8cm] {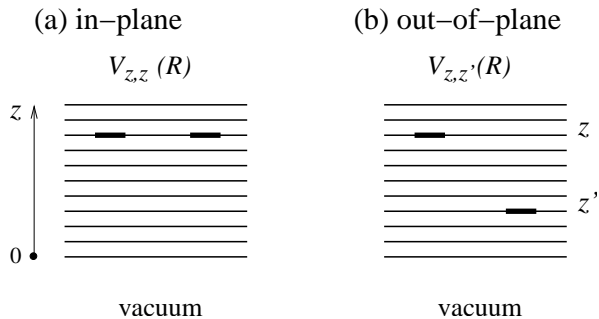}
   \caption{Geometry of the model studied in the
   paper. A layered superconductor fills the half space
   with $z>0$. The interaction between two pancake
   vortices is affected by the semi-infinite
   geometry and by the strong anisotropy, leading 
   to a strong logarithmic repulsion between pancake 
   vortices which reside within the same layer (a) or 
   to a weak logarithmic repulsion between pancake
   vortices in different layers (b).}
 \label{fig:model0}
\end{figure}

We consider a semi-infinite geometry with the
superconductor filling the upper half-space $z \ge 0$
and describe the vortex system within the London
theory (see Fig.\ \ref{fig:model0}). We model the system as a stack of 
two-dimensional superconducting layers of
thickness $d_\mathrm{s}$, separated by a distance $d$,
and with a penetration depth $\lambda_\mathrm{s}$. We
work in the limit of vanishingly small Josephson
coupling.  The basic topological defects are pancake
vortices\cite{clem91,clem04} which consist of vortices
with a two dimensional core limited to a single
superconducting layer.

The vortex interaction is mediated by currents set
up by the vortices via the Lorentz force. A finite sheet 
current ${\bf j}$ acts on the vortex core, producing a 
perpendicular force 
\begin{equation}\label{lorentz}
 {\bf F} =  (\Phi_0/c)\,{\bf z}\times{\bf j}.
\end{equation}
To obtain the force between two vortices, one needs 
the sheet supercurrent generated by any of the two at 
the location of the core of the other one; then, the interaction 
is found by integrating the force (\ref{lorentz}) on 
the radial coordinate back from infinite distance.
The circular current defining a vortex is induced by the $2\pi$ 
twist of the phase $\varphi$ of the complex
order parameter $\psi = |\psi| e^{i\varphi}$, 
\begin{equation}\label{curr}
  {\bf j} = - \frac{c d_\mathrm{s}}{4\pi \lambda_\mathrm{s}^2} 
  \Bigl(\frac{\Phi_0}{2\pi}\nabla  \varphi +{\bf A}\Bigr).
\end{equation}
The vector potential ${\bf A}$ screens the action of
the driving term $\nabla\varphi = - {\bf n}_z \times {\bf R}/R^2$,
annihilating its action when an entire flux quantum $\Phi_0$ is accumulated.
This is the case for Pearl vortices in a single superconducting film 
which trap an entire flux $\Phi_0$ at a distance  
$\lambda_{\rss eff}= \lambda_s^2/d$.
However, in the case of layered 
systems\cite{clem91,pudikov,mints2000,decol04}, the presence
of other layers reduces the ability of the vortex to bind a full 
flux quantum and hence the action persists to infinity.

A central quantity in the discussion of the vortex-vortex
interaction is the vector potential field ${\bf A}$ produced by a 
pancake vortex placed in a semi-infinite sample.
In the London approximation, a pancake vortex placed at $z'$ 
generates a vector potential satisfying the following
set of equations
\begin{eqnarray}
   &&\nabla^2 {\bf A}\!-\!\frac{1}{\lambda^2}{\bf A}
    = \frac{d}{\lambda^2}{\bf\nabla}\varphi\,\, \delta(z-z'),
    \quad z>0,
   \nonumber \\ 
   &&\nabla^2 {\bf A} = 0,
   \quad\quad\quad\quad  z<0.  \label{az}
\end{eqnarray}
Note that Eqs.\ (\ref{az}) are symmetric with respect to 
$z\leftrightarrow z'$. In (\ref{az}) we treat the layered 
system within a continuum approximation, 
neglecting small modulations in ${\bf A}$
(of subleading order $\sim$ $d/\lambda$) across the layers. 
Such a description is consistent
once the layer thickness $d_\mathrm{s}$ and the material
penetration depth $\lambda_\mathrm{s}$ are replaced by 
the inter-layer spacing $d$ and the bulk penetration depth 
$\lambda^2=\lambda_\mathrm{s}^2 d/d_\mathrm{s}$, respectively.
By solving Eq.\ (\ref{az}), we obtain the vector potential 
at the point $(R,z)$ inside
the superconductor produced by a pancake vortex placed at
the origin of the layer at $z'$,
\begin{align}\label{a_semi}
   A_{\phi}(R,z \geq 0,z' \geq 0) 
  =\frac{\Phi_0 d}{2\lambda^2}\int_0^{+\infty}
   \frac{dK}{2\pi}\frac{J_1(KR)}{K_+} \nonumber \\
   \qquad\times
    [f_{z-z'}(K) + \beta(K)f_{z+z'}(K)],
\end{align}
with $f_z(K) = \exp(-K_+|z|)$ and $\beta(K) = (K_+-K)/(K_+ + K)$
(due to the cylindrical symmetry of (\ref{az}), the vector potential 
has only an azymuthal component, i.e., along the direction of 
the unit vector ${\bf n}_\phi = {\bf n}_z\times{\bf R}/R^2$). 
A convenient quantity, which we will use in the following discussion, 
is the total flux $\Phi_\mathrm{t}(z,z')$ trapped in the layer at $z$ due to a 
pancake vortex placed at a distance $z'$ from the surface; 
Eq.\ (\ref{a_semi}) provides the result
\begin{equation}\label{phi_t_surf}
   \Phi_\mathrm{t}(z,z')=\frac{\Phi_0 d}{2\lambda}
   \Bigl(e^{-|z-z'|/\lambda}+e^{-(z+z')/\lambda}\Bigr).
\end{equation}
In the following, we will make use of these results to derive
the pancake vortex interaction between co-planar vortices 
(next subsection) and vortices in different layers (Sect.\ \ref{sec:oop}). 

\subsection{In-plane interaction}

We first consider two pancake vortices residing 
on the same layer, cf.\ Fig.\ \ref{fig:model0}(a). Their interaction
includes the contributions from both terms in Eq.\ (\ref{curr}), the 
driving source  $\nabla\varphi$ and the vector 
potential ${\bf A}$.
While the phase produces currents decreasing $\propto 1/R$, 
magnetic screening is effective at scales larger than
the penetration depth $\lambda$. To leading order, the 
vector potential can be written in
terms of the trapped magnetic flux $\Phi_\mathrm{t}(z)\equiv
\Phi_\mathrm{t}(z,z)$ 
which threads the layer where the vortex resides:
$A_\phi(R \gg \lambda,z,z) \approx \Phi_\mathrm{t}(z)/2\pi R$. 
Combining Eqs.\ (\ref{lorentz}) and  (\ref{curr}), 
we can extract the in-plane potential for two 
vortices at a distance $R$ by integrating 
over the radial coordinate from $\xi$, the coherence length, 
to $R$
\begin{equation} \label{vzz0}
  V_{z,z}(R) \approx {- 2\varepsilon_0d}\left\{
   \begin{array}{lr}
    {\ds \ln\frac{R}{\xi}}, & R\ll\lambda, \\
    \noalign{\vskip 5 pt}
    {\ds \ln\frac{\lambda}{\xi}
    +\Bigl[1-\frac{\Phi_\mathrm{t}(z)}{\Phi_0}\Bigr]
    \ln\frac{R}{\lambda}}, &
    \lambda\ll R,
   \end{array}\right.
\end{equation}
where $\Phi_\mathrm{t}(z)$ is the magnetic flux trapped
by a pancake vortex placed at a distance $z$ away from the surface.
From (\ref{phi_t_surf}), we find that the trapped flux $\Phi_\mathrm{t}(z)$
interpolates between the values $(d/\lambda)\Phi_0$ on the surface and
 $(d/2\lambda)\Phi_0$  in the bulk,
\begin{equation}\label{phi_t0}
  \Phi_\mathrm{t}(z) = \frac{\Phi_0 d}{2\lambda}
   \Bigl(1+e^{-2z/\lambda}\Bigr).
\end{equation}

However, we can safely ignore the small modifications of 
order $d/\lambda$ in the potential (\ref{vzz0}) 
arising from magnetic screening and 
assume an in-plane interaction which is independent of $z$. 
Within this approximation, vortices in the same layer feel 
a repulsive logarithmic interaction,
\begin{equation}\label{vzz}
  V_{z,z}(R) \approx - 2 \varepsilon_0 d \ln (R/\xi).
\end{equation}
This potential corresponds to the one between charges in the 
two-dimensional Coulomb gas, or, equivalently, the one-component 
plasma (OCP), with charge $e^2 \rightarrow 2\varepsilon_0 d$.

\subsection{Out-of-plane interaction}\label{sec:oop}

For two pancake vortices in different layers (cf. Fig.\ \ref{fig:model0}(b)), 
the current (\ref{curr}) which enters the Lorentz force (\ref{lorentz}) 
does not contain the contribution from the driving phase $\nabla \varphi$. Hence, the force in (\ref{curr}) is due to the vector potential $A$ alone.
The central quantity in the discussion is the vector field (\ref{a_semi}) which is 
produced by a single pancake vortex placed in a semi-infinite sample.
\begin{figure}[t]
\centering
   \includegraphics [width= 5.7cm] {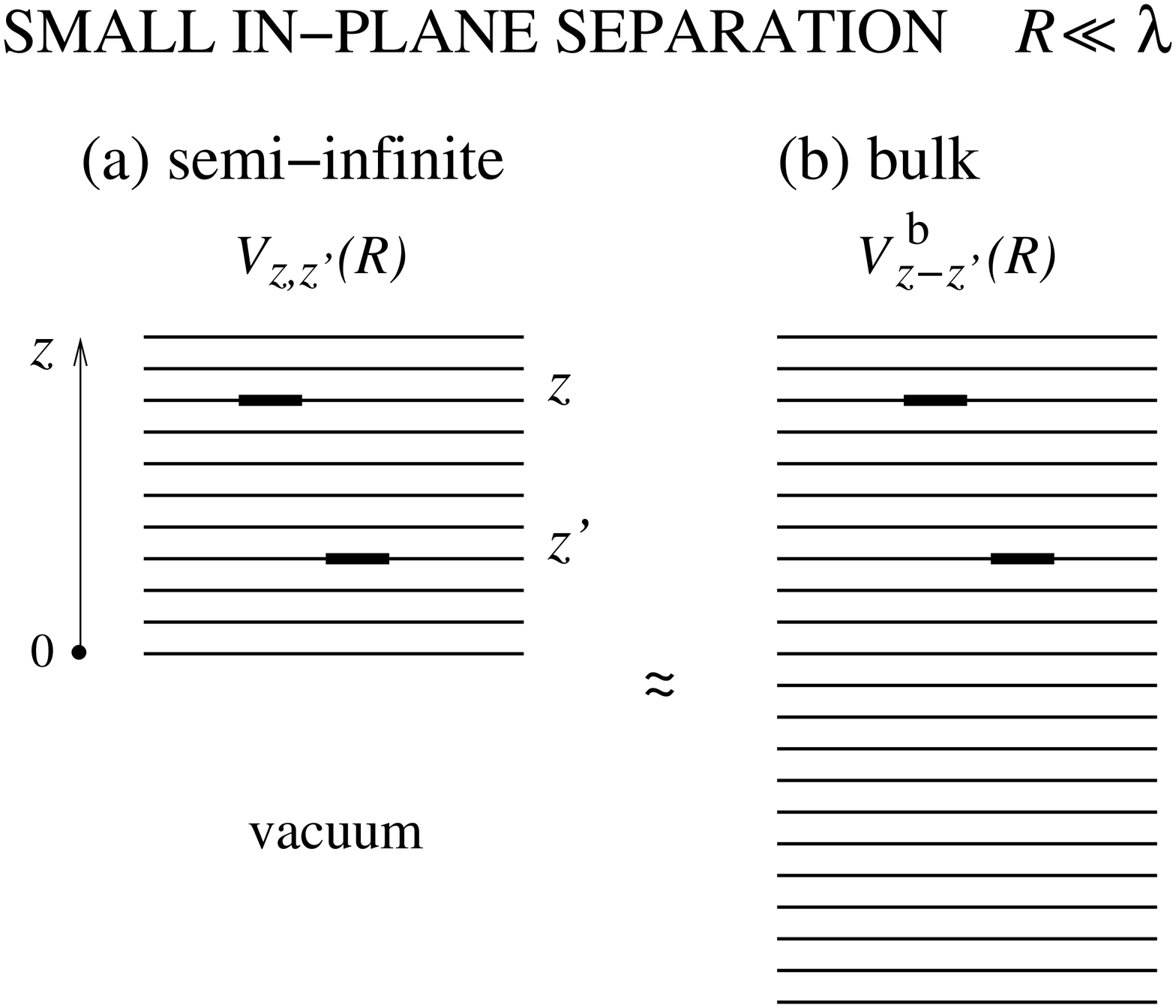}
   \caption{For small in-plane separation, i.e., 
   $R\ll \lambda$, 
   the total energy for the out-of-plane interaction 
   in a semi-infinite system (a) is equivalent
   to the one of a translation
   invariant bulk (b) system $V^\mathrm{b}_{z-z'}(R)$.}
 \label{fig:model}
\end{figure}

Inserting the vector potential (\ref{a_semi}) into
(\ref{lorentz}) and (\ref{curr}), we obtain an expression for
the force between two pancake vortices 
residing in different layers $z \neq z'$. The
integration over $R$ provides the out-of-plane interaction
\begin{eqnarray}\label{vzz1}
   \lefteqn{V_{z,z'}(R)
   = -\varepsilon_0 d^2
   \int_0^{+\infty}dK\frac{J_0(KR)}{K K_+}} \\
   &&\qquad \times [f_{z-z'}(K) + \beta(K)f_{z+z'}(K)].
   \nonumber 
\end{eqnarray}
This interaction is given by the sum of a bulk- (first) 
$V^\mathrm{b}_{z-z'}(R)$ 
and a stray-field term (second) $V^\mathrm{s}_{z,z'}(R)$, 
where the latter becomes negligible at a distance 
$\sim\lambda$ away from the surface.

For small in-plane distances, i.e., $R\ll \lambda$,
the contribution of the stray-field potential to
the overall vortex interaction energy can be neglected:
$V^\mathrm{s}_{z,z'}(R)\ll V^\mathrm{b}_{z-z'}(R)$.
Thus, for $R\ll\lambda$, the potential coincides 
with the bulk one\cite{giannireview} 
\begin{equation}\label{v_as_surf}
  V_{z,z'}(R) \approx \varepsilon_0d \frac{d}{\lambda}
    \left\{ \begin{array}{lr}
        {\ds \frac{R^2}{4|z-z'|\lambda},} 
        & R \ll |z-z'| \ll \lambda, \\
        \noalign{\vskip 5 pt}   
        {\ds \frac{R}{\lambda},} &
        |z-z'| \ll R \ll \lambda.
   \end{array}\right. 
\end{equation}
As expected, the out-of-plane interaction is smaller 
with respect to the in-plane expression (\ref{vzz}) 
due to the small pre-factor $d/\lambda$. However,
the out-of-plane interaction extends over many ($\lambda/d$) layers, 
underlying its importance. In the regime $R\ll\lambda$, 
the attractive interaction between a single pancake vortex 
and a semi-infinite stack then is reduced close to the surface 
due to the missing planes for $z<0$.

\begin{figure}[t]
\centering
   \includegraphics [width= 8.5cm] {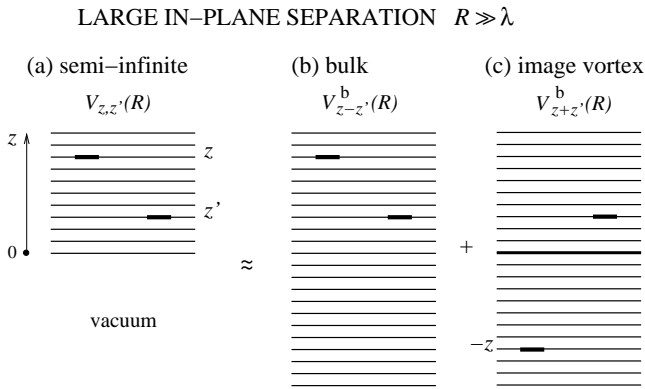}
   \caption{For large in-plane separation, i.e., 
   $R\gg \lambda$, 
   the total energy for the out-of-plane interaction 
   in a semi-infinite system (a)
   can be split in (b) a translation
   invariant bulk term $V^\mathrm{b}_{z-z'}(R)$ and (c) an 
   additional one $V^\mathrm{b}_{z+z'}(R)$ that can 
   be interpreted in terms 
   of an image vortex placed in $-z$. 
   The stray field contributes via an additional algebraic correction, 
   cf.\ (\ref{vzz'r})
    }
 \label{fig:model2}
\end{figure}

On the other hand, for large in-plane distances, i.e., 
$R\gg \lambda$, we find relevant modifications
of the out-of-plane interaction at the surface
\begin{align}\label{vzz'r0}
    V_{z,z'}(R) &\approx \varepsilon_0 d \frac{d}{\lambda} 
    e^{-|z-z'|/\lambda}\ln \frac{R}{\lambda} \nonumber \\
   & + \varepsilon_0 d \frac{d}{\lambda} e^{-(z+z')/\lambda}
    \Bigl(\ln \frac{R}{\lambda} + \frac{2\lambda}{R}\Bigr).    
\end{align}
The first term has a bulk origin, whereas
the terms which follow vanish away from
the surface. Combining terms, we can rewrite
(\ref{vzz'r0}) in the more convenient form
\begin{equation}\label{vzz'r}
   V_{z,z'}(R) = 2 \varepsilon_0 d\Bigl[ 
   \frac{\Phi_\mathrm{t}(z,z')}{\Phi_0}\ln \frac{R}{\lambda} + 
   \frac{d}{\lambda} e^{-(z+z')/\lambda} 
   \frac{\lambda}{R} \Bigr].
\end{equation}
We identify two different contributions: a logarithmic attractive 
interaction (first) and an algebraic repulsive potential (second).
The logarithmic interaction can be analyzed by considering the
presence of mirror vortices in $z<0$. The two terms in 
$\Phi_\mathrm{t}(z,z')$ can be viewed as the contributions of 
two bulk vortices, one at $z$ and a mirror one at $-z$, see 
Fig.\ \ref{fig:model2}. Hence, if we consider
only the logarithmic term in (\ref{vzz'r}), the interaction between
a pancake vortex and a stack in a semi-infinite system is
equivalent to that in a bulk system. Modifications appear
due to the stray magnetic field, generating the 
residual algebraic potential in (\ref{vzz'r}). 
This correction $\propto\lambda/ R$ is responsible for the  
weak algebraic interaction between vortex tips at large 
distances\cite{pearl2}. 

Summarizing, for short in-plane distances, $R\ll \lambda$, 
the strength of the out-of-plane potential is not affected by 
the semi-infinite geometry (see Fig.\ \ref{fig:model}) and 
the effect of the surface is limited to the lack of superconducting 
planes at $z<0$. On the other hand, for large in-plane 
distances, $R\gg \lambda$, the missing planes are compensated
by the contributions from the mirror term in (\ref{vzz'r0}), 
see Fig.\ \ref{fig:model2}. 
Surface effects then are due to the stray magnetic field generating
an additional algebraic repulsion, cf.\ Eq.\ (\ref{vzz'r}).

\section{DFT-substrate approach}

Here, we briefly discuss the DFT-substrate approach we will
use in our analysis (see Ref.\ \onlinecite{De-Col2006} for details). 
In the classical DFT one writes the total grand canonical free 
energy difference from the homogeneous liquid phase as a functional 
of the (averaged) density profile. In an anisotropic system such as 
the pancake vortex system, it is convenient to separate the in-plane 
dependence from the out-of-plane one. Hence, the DFT free energy
takes the form
\begin{align}
   \lefteqn{\frac{\delta \Omega [\rho_z({\bf R})]} {T}
   = \int \frac{dz}{d}\, d^2{\bf R} \, \Bigl[
   \rho_{z}({\bf R})\ln\frac{\rho_{z}({\bf R})}{\bar{\rho}}
   -\delta \rho_{z} ({\bf R})} 
   &
   \nonumber\\
  & & -\frac{1}{2}
   \int \frac{dz '}{d}\, d^2{\bf R'} 
   \delta \rho_{z}({\bf R})
   c_{z,z'}(|{\bf R}\!-\!{\bf R}'|)
   \delta \rho_{z'}({\bf R}')\Bigr], \label{func1}
\end{align}
where  $\rho_z({\bf R})$ is the averaged vortex density 
in the layer $z$ and $\delta \rho_z({\bf R}) = \rho_z({\bf R}) - \bar{\rho}$
is the deviation from the homogeneous liquid density 
$\bar{\rho}$. The only input needed in the DFT free energy
is the direct correlation function $c_{z,z'}(R)$.
In our approach we implement the substrate model\cite{dodgson00}
for the pair-correlation function by separating the contributions 
of the strong in-plane logarithmic repulsion from the weak 
out-of-plane but long-range attraction,
\begin{equation}\label{csub}
   c_{z,z'}(R) =  d c^{\rss 2D}(R) \delta(z-z') - V_{z,z'}(R)/T,
\end{equation}
where $V_{z,z'}(R)$ is the out-of-plane interaction of Eq.\ (\ref{vzz1}); 
the above approximation for the correlator is described in detail in Ref.\ 
\onlinecite{De-Col2006}. Within the planes, vortices are strongly 
correlated due to the repulsive logarithmic interactions (we neglect 
small contributions of order $d/\lambda$ and use (\ref{vzz})). 
Hence, we can approximate $c_{z,z}(R)$ with the direct correlation 
function $c^{\rss 2D}(R)$ of two-dimensional logarithmically 
interacting particles (also known as one-component plasma, OCP). 
Instead of using an approximate scheme, such as the hypernetted
chain equation\cite{menon96},  we use results of Monte Carlo simulations of the 
two-dimensional OCP at various coupling constants 
$\Gamma = 2\varepsilon_0 d/T$ to extract $c^{\rss 2D}(R)$. 
On the other hand, the out-of-plane direct correlator $c_{z,z'}(R)$ is 
determined within perturbation theory\cite{hansenbook} in terms 
of the weak out-of-plane potential; neglecting higher-order correlations, 
we approximate it with the leading unperturbed value $-V_{z,z'}(R)/T$.

At a mean-field level the thermodynamically stable state
corresponds to the minimal free energy configuration of the
functional (\ref{func1}). Then, the density functions
$\rho_z(R)$ must obey the saddle-point equation
\begin{equation}\label{min1}
  \ln \frac{\rho_z({\bf R})}{\bar{\rho}} = \int
  \frac{dz'}{d}\int d^2{\bf R'} \, c_{z,z'} (|{\bf
  R}\! -\! {\bf R'}|) \delta \rho_{z'}({\bf R'}).
\end{equation} 
A key quantity in our discussion is the molecular 
field\cite{ramakrishnan79,ramakrishnan82,chakrabarti94}
 $\xi_z({\bf R})$  defined through
\begin{equation}\label{molfield}
  \xi_z({\bf R}) = \ln [\rho_z({\bf R})/\bar{\rho}].
\end{equation}
Combining the saddle point equation (\ref{min1}) with (\ref{molfield}), 
the molecular field becomes
\begin{equation}\label{molfield2}
  \xi_z({\bf R}) =  \int \frac{dz'}{d}\int d^2{\bf R'} \, c_{z,z'}
  (|{\bf R}\! -\! {\bf R'}|)
  \delta \rho_{z'}({\bf R'})
\end{equation}
and represents the average potential produced by the modulated 
density.  However, whereas (\ref{molfield}) defines the molecular
field everywhere, Eq.\ (\ref{molfield2}) is only valid at the minimum.

Next, instead of seeking the exact form
$\rho_z({\bf R})$ which solves the non-linear integral
equations (\ref{min1}), we restrict our analysis to
a simple family of periodic functions which describes		
the modulations of the density in the solid phase.
In the following, we concentrate on the simplest case,
retaining only the first Fourier components of the
density in a triangular lattice,
\begin{equation}\label{ansatzdft}
    \frac{\rho_z({\bf R})}{\bar{\rho}} = 1 +
    \sum_{|{\bf K}_1|=G} \mu_z e^{i {\bf K}_1\cdot {\bf R}}
    = 1 +\mu_z g({\bf R}),
\end{equation} 
where the vectors ${\bf K}_1$
are the first reciprocal lattice vectors of the
frozen structure of length $G$ which is related to 
the lattice constant $a_{\scriptscriptstyle \triangle}$ 
via $G= 4\pi/\sqrt{3}a_{\scriptscriptstyle \triangle}$, 
and $\mu_z = \delta \rho_z(G)/\bar{\rho}$ is the Fourier 
component of the density with wave length $G$. 
In the following discussion, we consider 
the vortex system as incompressible and neglect the relative density 
change  $\delta \rho({\bf K} = 0)/\bar{\rho}$, hence the length of the
first reciprocal lattice vectors can be written as 
$G=(8\pi^2\bar\rho/\sqrt{3})^{1/2}$ (see Ref.\ \onlinecite{De-Col2006} 
for a discussion of how to include the effects of a finite compressibility). 
The function
$g({\bf R}) = \sum_{|{\bf K}_1|=G} 
   e^{i {\bf K}_1 \cdot {\bf R}}= 2 \cos (G x) 
   + 4 \cos(G x/2)\cos(\sqrt{3} G y/2)$
includes the sum over the six first reciprocal lattice vectors. 
We also write a similar Ansatz for the molecular field
\begin{equation}\label{ansatzxi}
  \xi_z({\bf R}) = \zeta_z +  \xi_z g({\bf R}),
\end{equation}
retaining only the zeroth and first Fourier
components, $\zeta_z$ and $\xi_z$ respectively. 
The Fourier components of $\rho({\bf R})$ and $\xi({\bf R})$
are not independent and can be related through (\ref{molfield}). 
Filtering out the zeroth and first Fourier components of $\rho_z({\bf R}) = 
\bar\rho(1 + \mu_z g({\bf R})) = \bar \rho \exp (\zeta_z + \xi_z g_{K_1}({\bf R}))$ 
and using
\begin{align}\label{orto1}
  \frac{1}{a}\int_a d^2{\bf R}g({\bf R}) = 0, \quad\quad
   \frac{1}{a}\int_a d^2{\bf R}[g({\bf R})]^2 = 6,
\end{align}
we obtain the dependencies
\begin{equation} \label{ximu}
  \zeta_z = -\Phi(\xi_z), \quad\quad \mu_z   = \Phi'(\xi_z)/6,
\end{equation}
where we have defined the function
\begin{equation}\label{phixi}
  \Phi(\xi) = \ln \left[ 
  \frac{1}{a}\int_a d^2{\bf R}\, e^{\xi g({\bf R})}
  \right].
\end{equation}
Inserting the above Ansatz for the order-parameter profile $\delta \rho$
into the free energy (\ref{func1}), we obtain the associated free energy 
density functional ($A$ is the sample area)
$\delta \omega/T = (1/\bar\rho A) \delta \Omega/T $
expressed through $\mu_z$
\begin{equation}
   \frac{\delta \omega[\mu_z]}{T}\! =\!\! 
   \int_0^\infty \!\!\frac{dz}{d}\Bigl[ 6 \xi_z \mu_z\!-\!\Phi(\xi_z)  
    -\! 3 \!\! \int_0^\infty\!\!\frac{dz'}{d} \bar{c}_{z,z'} \mu_{z'}\mu_z \Bigr], 
   \label{free20a}
\end{equation}
where
\begin{equation}\label{czz1}
 \bar{c}_{z,z'} = \bar{c}^{\rss 2D} d \delta(z-z') - V_{z,z'}(G)/T
\end{equation}
is the dimensionless Fourier transform\cite{note1}
of the direct correlation 
function $\bar{c}_{z,z'} = c_{z,z'}(G)$ evaluated at the first reciprocal 
lattice vector $G$. Consistently, we have defined 
$\bar{c}^{\rss 2D} = c^{\rss 2D}(G)$.
The effects of stray fields are accounted for within 
the out-of-plane interaction $V_{z,z'}(G)$ of Eq.\ (\ref{vzz1}),
\begin{align}
   \frac{V_{z,z'}(G)}{T} & = \!  
   - \frac{2\pi\bar\rho\Gamma d}{G^2 \lambda^2 G_+} 
   \Bigl( e^{-G_+|z-z'|} + \frac{G_+\! -\! G}{G_+\! +\! G}e^{-G_+(z+z')}\Bigr)
    \nonumber \\
   & = - \bar{\alpha} (\bar{f}_{z-z'} + \bar{\beta} \bar{f}_{z+z'}). 
   \label{vzz'def}
\end{align}
In (\ref{vzz'def}) we have defined $\bar{f}_z = f_z(G)= \exp(-G_+ |z|)$, 
$\bar{\alpha} = \alpha(G) = 2\pi \bar \rho \Gamma d/G^2 \lambda^2 G_+$, 
$\bar{\beta} =(G_+ - G)/(G_+ + G)$, and $G_+=\sqrt{G^2+1/\lambda^2}$. 
Note the additional factor $\bar\rho$ due to the dimensionless definition 
of the Fourier transform. Combining (\ref{free20a}) with (\ref{czz1}) 
and (\ref{vzz'def}) and after few simple algebraic manipulations we 
obtain the functional
\begin{align}
   \frac{\delta \omega[\mu_z]}{T} & = \int_0^{\infty} 
   \frac{dz}{d}\Bigl[ 
   \frac{\delta \omega^{\rss 2D}_{\rss sub} (\mu_z)}{T}
   \nonumber \\ 
   & + \frac{3\bar{\alpha}}{2}\int_0^{\infty} 
   \frac{dz'}{d} (\bar{f}_{z-z'} + \bar{f}_{z+z'})
   (\mu_{z} - \mu_{z'})^2 \Bigr] \nonumber  \\
   & + \frac{6 \bar{\alpha} G}{G + G_+} \int_0^\infty 
   \frac{dz}{d}\int_0^\infty \frac{dz'}{d}\,
   \bar{f}_{z+z'}\mu_z\mu_{z'}. \label{free2a}
\end{align}
The first term in (\ref{free2a}) describes the local two-dimensional 
free energy of a uniform system
\begin{equation}
  \delta \omega^{\rss 2D}_{\rss sub}(\mu)/T
    = 6\xi(\mu) \mu - \Phi(\xi(\mu))  - 3\bar{c}^{\rss 2D}_{\rss sub}  \mu^2,
    \label{free2d0}
\end{equation}
where $\xi$ has to be understood as a function of $\mu$ through (\ref{ximu}). 
In (\ref{free2d0}) we have defined the correlator 
\begin{equation}\label{c3d_s}
   \bar c^{\rss 2D}_{\rss sub}\! =  
   \bar c^{\rss 2D} -  V_{\rss stack} (G)/T,
\end{equation}
where the out-of-plane interactions contribute to the correlator 
through the total stack potential 
\begin{align}\label{vstackg}
   -\frac{V_{\rss stack}(G)}{T} &
    = -\bar\alpha\int_{0}^{+\infty} \!\!
   \frac{dz}{d} (\bar{f}_{z-z'} + \bar{f}_{z+z'})
    =  \frac{2 \bar\alpha}{dG_+}.
\end{align}
The additional non-local terms in (\ref{free2a}) quantify the energy cost due 
to variations of the order parameter (the non-local `elastic' term, second line) 
and the effect of the surface (third line). 

The free-energy (\ref{free2a}) has to be compared with the phenomenological 
Landau-type functional of Eq.\ (\ref{landau_s}). Both expressions exhibit
a similar structure composed by a sum of bulk, `elastic',
and surface terms. However, the analysis of the free energy (\ref{free2a}) 
is complicated by the non-local nature of the `elastic' and surface terms. 

\section{Bulk System}

Before concentrating on the surface problem, we briefly review
the case of an infinite bulk system. The results in this section will 
be the basis for the discussion of the melting transition in the 
semi-infinite geometry which we will present in the following sections. 

In a bulk system, we can drop the surface term and the free energy 
(\ref{free2a}) becomes
\begin{align}   
   \frac{\delta \omega[\mu_z]}{T} &  =  \int_{-\infty}^{\infty} \frac{dz}{d}\Bigl[ 
   \frac{\delta \omega^{\rss 2D}_{\rss sub} (\mu_z)}{T} \nonumber \\
   &\quad +  \frac{3\bar \alpha}{2}\int_{-\infty}^{\infty} 
   \frac{dz'}{d} \bar{f}_{z-z'}
   (\mu_{z'} - \mu_z)^2 \Bigr]. \label{free2abulk}
\end{align}
Below, we first concentrate on the thermodynamically stable state 
(see also Ref.\ \onlinecite{De-Col2006}) and then analyze 
the properties of a solid-liquid interface.

\subsection{Thermodynamic phase diagram}\label{sec:melting_in}

In thermodynamic equilibrium, all superconducting 
layers are equivalent and the order parameter $\mu_z$ 
becomes independent of the layer position $z$, hence we 
write $\mu_z = \mu$. For a constant order-parameter profile the 
second term in (\ref{free2abulk}) is zero and only the bulk 
free energy of Eq.\ (\ref{free2d0}) is relevant. 
The key quantity is the effective three-dimensional
correlator $\bar{c}^{\rss 2D}_{\rss sub}$ which is given by the sum 
of two contributions: the OCP term $\bar{c}^{\rss 2D}$ 
and the stack potential $V_{\rss stack}(G)$ of Eq.\ (\ref{vstackg}). 
Whereas the first depends only on the  temperature, the latter 
depends also on the vortex density and thus on the magnetic field,
\begin{align}\label{c2dbt}
  \bar{c}^{\rss 2D}_{\rss sub}(T,B) & =  \bar{c}^{\rss 2D}(T) 
     + \frac{4\pi\bar{\rho}\varepsilon_0d}{TG^2}\frac{1}{1+\lambda^2 G^2}
    \\
  & = \bar{c}^{\rss 2D}(T) + \frac{\sqrt{3}\varepsilon_0d}{2 \pi T} 
  \frac{1}{[1+ (8\pi^2/\sqrt{3})B/B_\lambda] },  \nonumber
\end{align}
where we use $\bar{\rho}/G^2 = \sqrt{3}/(8\pi^2)$ and 
$\lambda^2 G^2= (8\pi^2/\sqrt{3})
B/B_\lambda$, with $B_\lambda = \Phi_0 /\lambda^2$.

The temperature and magnetic field dependencies entering 
$\bar{c}^{\rss 2D}_{\rss sub}(T,B)$ change the coefficient 
of the quadratic term as in the $\phi^4$ Landau theory. 
As a function of $\mu$, the free energy exhibits the shape of a 
Landau theory describing a first-order phase transition 
(this function is plotted in Ref.\ \onlinecite{De-Col2006}): for small
values of the correlator $\bar{c}^{\rss 2D}_{\rss sub}$, corresponding 
to large temperatures and fields, $\delta \omega^{\rss 2D}(\mu)$ 
exhibits only one minimum at $\mu_\mathrm{l} = 0$ with a value 
$\delta \omega^{\rss 2D}(0)/T = 0$, corresponding to a (homogeneous) 
liquid phase. Decreasing the temperature and/or field 
(which corresponds to increasing values of 
$\bar{c}^{\rss 2D}_{\rss sub}$), a second local minimum 
at $\mu_\mathrm{s}>0$ (metastable solid) appears in addition 
to the liquid one at $\mu_\mathrm{l}=0$.
Freezing occurs when the liquid and solid minima assume 
the same value of the free energy, i.e., when 
$\delta \omega^{\rss 2D}(\mu_\mathrm{s}) = 0$.
This condition is equivalent to a simple equation for the 
correlator \cite{ramakrishnan82}
\begin{equation}\label{critical_inc}
  \bar{c}^{\rss 2D}_{\rss sub}(T,B) = \bar{c}_c \approx 0.856.
\end{equation}
Solving this equation for $B$ yields the melting line, see Ref.\ \onlinecite{De-Col2006}.
Going to even lower temperatures, $\bar{c}^{\rss 2D}$ further 
increases, the solid minimum decreases further, 
$\delta \omega^{\rss 2D}(\mu_\mathrm{s})/T < 0$,
and the crystal becomes the only thermodynamically stable phase.

\subsection{Solid-liquid interface}\label{sec:sol-liq}

Next, we consider a solid-liquid interface in an infinite system, for
which both terms in (\ref{free2abulk}) are important.
Along the melting line $B_\mathrm{m}(T)$ the solid and the 
liquid assume the same value of the free energy and can 
coexist. Here, we analyze the situation where half of the system 
is in the liquid state and the other half is solid. The corresponding 
profile of $\mu_z$ takes the form of a soliton with boundary 
conditions $\mu_{z\rightarrow -\infty} = 0$ (liquid) and  
$\mu_{z\rightarrow +\infty} = \mu_\mathrm{s}\approx 0.51$ (solid), 
where $\delta \omega^{\rss 2D}_{\rss sub}(0)/{T}  = 
\delta \omega^{\rss 2D}_{\rss sub}(\mu_\mathrm{s})/T = 0$. 
The properties of the interface between the two phases 
depend on the non-local term in the free-energy 
functional (\ref{free2abulk}). For a large part 
of the phase diagram (see Appendix\ \ref{app:ge}), it is 
possible to approximate the full non-local theory with a local
one by proceeding with a gradient expansion
\cite{Haymet1981,Oxtoby1982} of the kernel in (\ref{free2abulk}); 
inserting $\mu_{z'}\approx \mu_z+(d\mu_z/dz)(z'-z)$ into 
(\ref{free2abulk}), we obtain 
\begin{equation}
    \frac{\delta \omega[\mu_z]}{T} = \int_{-\infty}^{\infty} \frac{dz}{d}   
    \Bigl[\frac{\delta \omega^{\rss 2D}_{\rss sub}(\mu_z)}{T}+\frac{\ell^2}{2}
    \Bigl(\frac{d\mu_z}{dz}\Bigr)^2\Bigr], 
    \label{freebulkgradient}
\end{equation}
with the elastic scale
\begin{align}
	\ell^2  & =  \frac{ 3 \bar\alpha}{d} \int_{-\infty}^{+\infty} 
	 dz\,  \bar{f}_{z} z^2 
	= \frac{12 \bar{\alpha}}{dG_+^3} = 
	 \nonumber \\
	& = 
	\lambda^2\, \frac{3\sqrt{3}}{2\pi}
	\frac{\Gamma}{[1+ (8\pi^2/\sqrt{3})
	B_\mathrm{m}(T)/B_\lambda]^2}.\label{ell0}
\end{align}
This local approximation describes well the full non-local free
energy if the profile $\mu_z$ varies slowly over the extension 
$1/G_+$ of the kernel. We have checked that this condition is 
fulfilled for $T \gtrsim 0.04\, \varepsilon_0 d$, which corresponds to
small and moderate magnetic fields $B \lesssim 0.5 B_\lambda$, 
by comparing the shape of the soliton in the non-local and local theories
(see Appendix\ \ref{app:ge}).

\begin{figure}[t]
\centering
   \includegraphics [width= 6.8 cm] {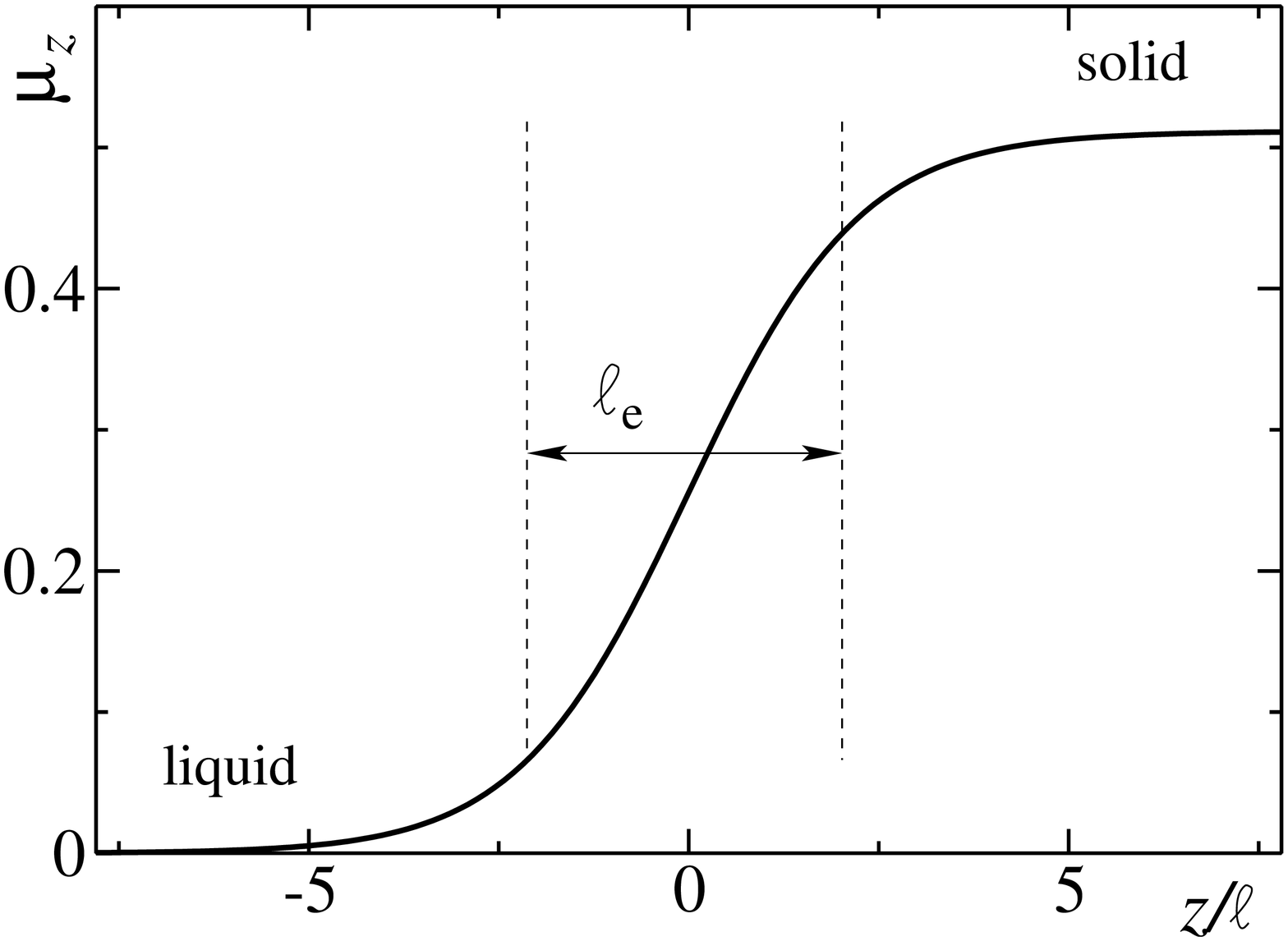}
   \caption
   [Solid-liquid interface]
   {Profile of the order parameter $\mu_z$ describing an interface
   between a liquid ($\mu_{z\rightarrow -\infty}=\mu_\mathrm{l}=0$) 
   and a solid ($\mu_{z\rightarrow +\infty} = \mu_\mathrm{s}
   \approx 0.51$). The shape has been obtained numerically by solving the 
   differential equation (\ref{eqmotion0}). The width of the interface is 
   approximately $\ell_\mathrm{e} \approx 4.05\, \ell$ (we define it as the 
   full width at half maximum of the function $d\mu_z/dz$), where 
   $\ell$ is defined in (\ref{ell0}).}   
   \label{fig:interface}
\end{figure}

We can easily estimate the main properties of the 
solid-liquid interface, e.g.\ its width and free energy cost.
At small magnetic fields, we use the gradient expansion and
write the Euler-Lagrange equation of the free-energy functional 
(\ref{freebulkgradient})
\begin{equation}
  	\ell^2\frac{d^2\mu_z}{dz^2} = 
	\frac{d}{d\mu_z}
	\Bigl( 
	\frac{\delta \omega^{\rss 2D}_{\rss sub}(\mu_z)}{T}
	\Bigr)\label{eqmotion0}
\end{equation}
and the corresponding energy conservation (the integration 
constant is zero for a soliton solution)
\begin{equation}\label{ensol0}
	\frac{d\mu_{z}}{dz}= \frac{1}{\ell}
	\sqrt{\frac{2 \delta \omega^{\rss 2D}_{\rss sub}(\mu_z)}{T}}.
\end{equation}
The energy $E_\mathrm{sl}$ of the interface (soliton) is given by 
\begin{align}
  E_\mathrm{sl}  & = 
  T \int \frac{dz}{d} \ell^2 \Bigl(\frac{d\mu_z}{dz}\Bigr)^2 
  = T \frac{\ell}{d} \int_0^{\mu_\mathrm{s}} 
  d\mu \sqrt{\frac{2\delta\omega^{\rss 2D}_{\rss sub}(\mu)}{T}}
 \label{esl0} \\
   & =
  C T \frac{\ell}{d}\mu_\mathrm{s}\sqrt{
  \frac{2\delta\omega^{\rss 2D}_{\rss max}}{T}}, \nonumber
\end{align}
where $\delta \omega^{\rss 2D}_{\rss max} 
\approx 0.0065\, T$ is the barrier in the energy density between the solid 
and liquid minima on the melting line and $\mu_\mathrm{s} \approx 0.51$. 
The constant $C$ is of order unity and requires to evaluate the integral 
in (\ref{esl0}): a numerical calculation gives $C  \approx 0.69$. 

Next, we determine the width of the soliton $\ell_\mathrm{e}$,
which we define as the full width at half maximum of the derivative 
$d\mu_z/dz$. Approximating $d\mu_z/dz \approx \mu_\mathrm{s}/
\ell_\mathrm{e}$ in (\ref{esl0}), we obtain an estimate 
for the soliton energy  $E_\mathrm{sl} \approx T\ell^2
(\ell_\mathrm{e}/d)(\mu_\mathrm{s}/\ell_\mathrm{e})^2$.
Combining this formula together with (\ref{esl0}), we find
\begin{equation}\label{elle1}
 \ell_\mathrm{e} \approx \frac{\mu_\mathrm{s}}
 {\sqrt{2\delta\omega^{\rss 2D}_{\rss max}/T}}\, \ell.
\end{equation}
From (\ref{elle1}), we obtain $\ell_\mathrm{e} \approx 6.33\,\ell$; 
however, a precise determination of the proportionality factor requires  
a numerical study, which yields $\ell_\mathrm{e} \approx 4.05\,\ell$, 
see Fig.\ \ref{fig:interface}. At low fields, $B \to 0$ 
(and $T \to T_{\rss BKT}$), 
the soliton becomes wider than the bulk penetration depth, 
$\ell_\mathrm{e} \approx 8 \lambda$, see (\ref{elle1}) and (\ref{ell0}). 
This confirms the validity of the gradient expansion for $B\to 0$, 
since in this limit 
$1/G_+\approx \lambda$ and $\ell_\mathrm{e} G_+ \approx 8$.

On the other hand, for large $B$, the gradient expansion
is not valid and we cannot use Eq.\ (\ref{elle1}). 
However, we have determined numerically the shape of the soliton
(see Appendix\ \ref{app:ge}) using the full non-local theory 
(\ref{free2abulk}) and found its width to decrease with increasing $B$
(the full theory produces a sharper interface than the one originating
from the gradient approximation).  
We can interpret this results in terms of the intra-layer interactions 
dominating over the inter-layer ones, leading to a system of decoupled 2D 
planes which melt independently.

\section{Surface melting}\label{sec:dft_surf}

The presence of a surface modifies the bulk free energy functional 
(\ref{free2abulk}) in two ways as it is clear from (\ref{free2a}): 
{\it i)} the superconductor occupies only a half-infinite space $z>0$, 
{\it ii)} stray magnetic fields modify the vortex interaction and hence 
the direct correlation function $\bar{c}_{z,z'}$ near the surface. 

The configuration of the system is derived from the saddle-point equation 
of (\ref{free2a}). Minimization with respect to $\mu_z$ provides us with the 
integral equation
\begin{eqnarray}
   &&\frac{\partial_{\mu_z} \delta \omega^{\rm
   \scriptscriptstyle 2D}_{\rss sub} (\mu_z)}{T}
   + 6\bar{\alpha}\!\! \int_0^\infty\! \frac{dz'}{d}
   [\bar{f}_{z-z'}+\bar{f}_{z+z'}] (\mu_z-\mu_{z'}) \nonumber\\
   &&\qquad\qquad\quad
   + \frac{12\bar{\alpha} G}{G+G_{\scriptscriptstyle +}}\! \int_0^\infty\! 
   \frac{dz'}{d}\, \bar{f}_{z+z'}\, \mu_{z'}= 0.
   \label{eqmotion_0}
\end{eqnarray}
The surface produces two additional terms when compared 
to an infinite system: a mirror term ($\propto\bar{f}_{z+z'}$) 
in the non-local elastic force and a pure surface term (second line), cf.\ 
the discussion of the out-of-plane interaction in Sec.\ \ref{sec:oop}. 

In the following, we first acquaint ourselves with the formalism by 
studying the $B\approx 0$ and large field limits. Then, we present 
the full analysis at low but finite values of the magnetic field. 

\subsection{Large magnetic field and $B\approx 0$ limits}\label{sec:limbb}

Let us consider first large magnetic fields, $B\gg B_\lambda$.
In this limit, the contribution of the out-of-plane
potential vanishes ($G\approx (B/\Phi_0)^{1/2} \to \infty$ and 
$\bar \alpha \to 0$) and the system decouples into independent 
two-dimensional systems. Only the intra-layer interaction remains 
relevant in this limit and the free energy is simply given 
by the bulk local contribution (\ref{free2d0}). The saddle point equation 
\begin{equation}\label{se_0}
  \partial_{\mu_z} \delta 
\omega^{\rm \scriptscriptstyle 2D}_{\rss sub} (\mu_z)/T=0
\end{equation}
does not depend on $z$ and the solution is given by a constant 
order parameter profile. Thus, no modifications occur at the surface 
for $B\gg B_\lambda$.

In the opposite limit of $B\to 0$ ($G\to 0$), $G_+\approx 1/\lambda$ 
and $\bar \alpha$ remain finite. The surface term in (\ref{free2a}) 
is negligible and the saddle point equation 
\begin{eqnarray*}
   \frac{\partial_{\mu_z} \delta \omega^{\rm
   \scriptscriptstyle 2D}_{\rss sub} (\mu_z)}{T}
   + 6\bar{\alpha}\!\! \int_0^\infty\! \frac{dz'}{d}
   [\bar{f}_{z-z'}\!+\!\bar{f}_{z+z'}] (\mu_z-\mu_{z'}) =0,
\end{eqnarray*}
contains an additional term as compared to (\ref{se_0}). 
However, we can render the problem translational invariant on the 
whole $z$-axis, by introducing mirror vortices in $z<0$ via 
$\mu_{-z} = \mu_z$ and changing $z'\to -z'$ in the integral of the mirror term. 
As a result, we can map the semi-infinite system back to the infinite 
bulk and the constant bulk solution remains valid also in this limit.
We can conclude that for both $B\approx 0$ and $B\gg B_\lambda$
the surface does not lead to a modification of the bulk behavior.

\subsection{Low magnetic fields}\label{sec:dftsurf}

At finite magnetic fields, all three terms of the saddle-point 
equation (\ref{eqmotion_0}) are relevant. To understand the 
effect of the surface term (third), it is convenient to look at the 
associated free energy expression (cf.\ (\ref{free2a})), which 
can be written as a weighted square of $\mu_z$,
\begin{align}
   \frac{\delta \omega^\mathrm{s}[\mu_z]}{T}  
    =  \frac{6\bar{\alpha} G}{G+G_{\scriptscriptstyle +}}
   \Bigl( \int_0^{\infty}\frac{dz}{d}\, e^{-G_+ z} \mu_z \Bigr)^2.\label{free_st}
\end{align}
This term favors the appearance of the liquid on the surface, 
similarly to the local surface term in (\ref{landau_s}). In real space, 
(\ref{free_st}) is associated with the $\propto 1/R$ repulsive potential 
induced by the stray magnetic field, cf.\ the second term in (\ref{vzz'r}).
In analogy to the Landau-theory description of Eq.\ (\ref{landau_s}),
this DFT functional can yield two different surface melting scenarios:
discontinuous, $O_1$, and continuous, $O_2$ (see introduction).

To make progress analytically, we have to simplify the non-local terms
in (\ref{eqmotion_0}). Concentrating on the bulk, i.e., $z\gg 1/G_+$, 
both mirror and surface terms can be ignored. For not too large values
of the magnetic fields, we can adopt a gradient expansion of 
the non-local elastic term $\propto \bar{f}_{z-z'}$, cf.\ Sec.\ \ref{sec:sol-liq} 
and Appendix\ \ref{app:ge}. As a result we obtain the differential
equation
\begin{equation}
   \ell^2 \frac{d^2\mu_z}{dz^2}
   = \frac{d}{d\mu_z} \frac{\delta \omega^{\rm\scriptscriptstyle
   2D}_{\rss sub}(\mu_z)}{T}
   \label{eqmotion}
\end{equation}
with $\ell^2 = (6\bar\alpha/d)\int_0^{+\infty} dz $ $\bar{f}_z z^2 = 12
\bar\alpha/d\, G_{\scriptscriptstyle +}^3$, cf. Eq.\ (\ref{ell0}). Equation 
(\ref{eqmotion}) has to be completed with a boundary condition, which 
has to be provided by the surface term in the integral equation. 
In the following, we restrict the analysis to small values of the order 
parameter at the surface in order to study the boundary 
between continuous and discontinuous surface melting scenarios. 
In this regime the bulk potential (\ref{free2d0}) can be approximated as
\begin{equation}\label{linf2d}
     \delta \omega^{\rm \scriptscriptstyle 2D}_{\rss sub}(\mu)/T
     \approx 3 (1-\bar{c}_c)\mu^2,
\end{equation}
where we used $\Phi(\xi)\approx 3\xi^2$ (cf.\ (\ref{phixi})), 
$\mu = \Phi'(\xi)/6\approx \xi$ (using (\ref{ximu}) together 
with (\ref{orto1})) and the value of the critical correlator 
$\bar{c}^{\rss 2D}_{\rss sub} = 
\bar{c}_c \approx 0.856$ at melting. The saddle point equation 
takes the form of a linear integral equation
\begin{align}
     6(1-\bar{c}_c)\mu_z &+ 6\bar{\alpha}\!\! \int_0^\infty\! \frac{dz'}{d}
   [\bar{f}_{z-z'}+\bar{f}_{z+z'}] (\mu_z-\mu_{z'})+ \nonumber \\
   &+ \frac{12\bar{\alpha} G}{G+G_{\scriptscriptstyle +}}\! \int_0^\infty\! 
   \frac{dz'}{d}\, \bar{f}_{z+z'}\, \mu_{z'}= 0
   \label{eqmotion_-1}
\end{align}
with three terms: a potential force (first), a non-local 
elastic force (second) and the drive at the surface (third). 
Separating the non-local terms from the local ones, we obtain
\begin{equation}
  \Bigl[6(1-\bar{c}_c) + \frac{12 \bar \alpha}{dG_+}\Bigr] \mu_z = 
  6 \bar \alpha \int_0^\infty \frac{dz'}{d}
  [\bar{f}_{z-z'}+\bar{\beta} \bar{f}_{z+z'}]\, \mu_{z'},\label{basta}
\end{equation}
where $\bar\beta$ was defined in (\ref{vzz'def}).
The second term in the left hand side carries the
normalization of the kernel $\bar{f}_{z}$ (from the integral 
$6\bar\alpha\int_0^\infty (dz'/d)[\bar{f}_{z-z'}+ \bar{f}_{z+z'}] = 
12\bar\alpha/G_+d$ in the elastic term of (\ref{eqmotion_-1})). 
Equation\ (\ref{basta}) can be  rewritten as
\begin{equation}
   \mu_z = \frac{G_{\scriptscriptstyle +}}{2(1+r)} \int_0^\infty\! dz'\,
   [\bar{f}_{z-z'}+\bar{\beta} \bar{f}_{z+z'}]\, \mu_{z'},
   \label{inteq}
\end{equation}
with $r=d G_{\scriptscriptstyle +}(1-\bar{c}_c)/2\bar\alpha = 
6(1-\bar{c}_c)/(G_+\ell)^2$. 

This kind of integral equation is commonly found in
the study of boundary problems, e.g. in the analysis
of surface effects on the superconducting transition
\cite{degennesbook,Mineev1999}. For the latter case,
the equation which determines the superconducting gap
$\Delta(r)$ is a non-local integral equation with a
structure similar to (\ref{basta}) (or (\ref{inteq})).
Close to the superconducting transition $T_c$, a
gradient expansion reduces the non-local equation
into the local Ginzburg-Landau equation. The presence
of the surface enters the kernel of the non-local theory 
via an additional contribution, corresponding
to the term $\propto \bar{f}_{z+z'}$ in (\ref{basta}).
However, a major difference between the two problems
is given by the different nature of the bulk phase
transitions, first order for the melting and second order
for the superconducting transition.  In the case of a
second-order phase transition the coefficient of the
quadratic term in the bulk free energy goes to zero at
the transition point, cf.\ the coefficient $\alpha \sim
(T- T_c)$ in the Ginzburg-Landau equation. Hence, the
term corresponding to the force $\propto6(1-\bar{c}_c)$ 
in the LHS of (\ref{basta}) is absent. In Eq.\ (\ref{inteq}) 
the difference between a first- and a second-order bulk 
transition enters in the normalization of the integral in the 
RHS, through the parameter $r$, which is $r\neq0$ for a first-order
transition and $r=0$ for a second-order one.

In general, the solution of an integral equation of the type (\ref{inteq}) 
is a non-trivial task, which usually cannot be carried out exactly. 
However, in the present case a straightforward solution is possible 
due to the particular exponential form of the kernel. In fact, taking the 
second derivative of (\ref{inteq}) and making use of the identity
\begin{equation}
   \frac{d^2}{dz^2}  e^{-G_+ |z-z'|} = -2 G_+ \delta (z-z') 
   + G_+^2 e^{-G_+ |z-z'|},
\end{equation}
we find that the integral equation yields the {\it same} differential equation 
(\ref{eqmotion}) as derived previously in the bulk by means of a 
gradient expansion, 
\begin{align}
  (1+r)\ell^2 \frac{d^2\mu_z}{dz^2} \approx \ell^2 \frac{d^2\mu_z}{dz^2}
  & = r (\ell G_+)^2 \mu_z \nonumber \\
  & = 6(1-\bar{c}_c) \mu_z; \label{eq_nlt}
\end{align}
here, we have used the linearized force 
$\partial_\mu \delta\omega^{\rss 2D}_{\rss sub}(\mu)/T = 6(1-\bar{c}_c) \mu_z$ 
and we can drop the small renormalization factor $(1+r)\approx 1$ 
in the LHS of the equation\cite{note2}. The differential equation (\ref{eq_nlt}) 
admits two {\it exponential} solutions. 
This has to be compared to the case of a second-order bulk 
transition, where, due to the absence of a linear term in the 
bulk free energy, one obtains the differential equation 
$d^2\mu_z/dz^2=0$ which is solved by a {\it linear} function 
\cite{degennesbook}. In both cases, the integral equation (\ref{inteq}) is 
equivalent to the {\it bulk} differential equation (\ref{eq_nlt}) and the
bulk solution goes through the surface region. Hence, the effect 
of the surface terms $\propto \bar\beta$ is only to provide a boundary 
condition at $z=0$ but does not modify the bulk solution. 
This lucky coincidence is due to the particular exponential structure of the 
kernel. Usually, the bulk solution is not valid in the vicinity of the 
surface and the problem becomes much more difficult to solve.

Next, we derive the boundary condition which is provided by 
the surface term. We first write the general solutions of the 
bulk differential equation
\begin{equation}
 \mu_z = Ae^{\gamma G_{\scriptscriptstyle +}z} + Be^{-\gamma 
 G_{\scriptscriptstyle +}z},
\end{equation}
where $\gamma^2 = r/(1+r)$; the boundary condition then follows 
from the ratio $A/B$. Here, we retain the small correction $(1+r)$
in (\ref{eq_nlt}) as we base our analysis on the linearized integral 
equation (dropping this term leads to results which are correct to 
order $r$, in agreement with the precision of the gradient expansion). 
Inserting this Ansatz back into (\ref{inteq}), we obtain the condition 
\begin{displaymath}
 A \Bigl(\frac{\bar\beta}{1-\gamma} - 
 \frac{1}{1+\gamma}\Bigr) 
 + B \Bigl(\frac{\bar\beta}{1+\gamma} - \frac{1}{1-\gamma}\Bigr)=0.
 \label{int_eq_ans}
\end{displaymath}
This requirement selects a unique value of the ratio $B/A$,
\begin{equation}
   \frac{B}{A} = \frac{\gamma(\bar\beta +1) + (\bar\beta -1)  }
   {\gamma(\bar{\beta} + 1) - (\bar{\beta} -1)}.
\end{equation}
Finally, we can calculate the logarithmic derivative at $z=0$, which 
gives the relevant boundary condition for our forthcoming analysis, 
cf.\ Eq.\ (\ref{bc_land}), and we find the result 
(we define $\mu_z^\prime = \partial_z \mu_z$) 
\begin{equation}
   \frac{\mu_z^\prime}{\mu_z}\Bigr|_{z=0}= 
   G_+\gamma \frac{A-B}{A+B}
   = G_{\scriptscriptstyle +}
   \frac{1-\bar{\beta}}{1+\bar{\beta}}= G.
   \label{bc}
\end{equation}
This final relation can also be derived directly without solving the differential
equation at the surface: calculating the value of the order 
parameter and its derivative at $z=0$ from (\ref{inteq}), one finds
\begin{align}
 & \mu_0  =  \frac{G_{\scriptscriptstyle +}}{2(1+r)} 
 (1+\bar\beta) \int_0^{\infty} dz \bar{f}_z\mu_z, \nonumber \\
 &\frac{d\mu_z}{dz}\Bigr|_{z=0} =  \frac{G_{\scriptscriptstyle +}}{2(1+r)} 
 G_{\scriptscriptstyle +}(1-\bar\beta) \int_0^{\infty} dz \bar{f}_z\mu_z \nonumber.
\end{align}
The ratio $\mu'_0/\mu_0$ then does not depend on the function 
$\mu_z$ which has dropped out. As a result we recover (\ref{bc}), which
also remains valid for the case of a bulk second-order phase 
transition ($r=0$).  In the limit $B\approx 0$, the boundary condition 
becomes $\mu^{\prime}_{0}=0$, since $G \approx 0$. 
Hence, the constant bulk solution goes through and the surface is 
not relevant, in agreement with the results of Sec.\ \ref{sec:limbb}.

\begin{figure}[t]
\centering
    \includegraphics [width= 8 cm] {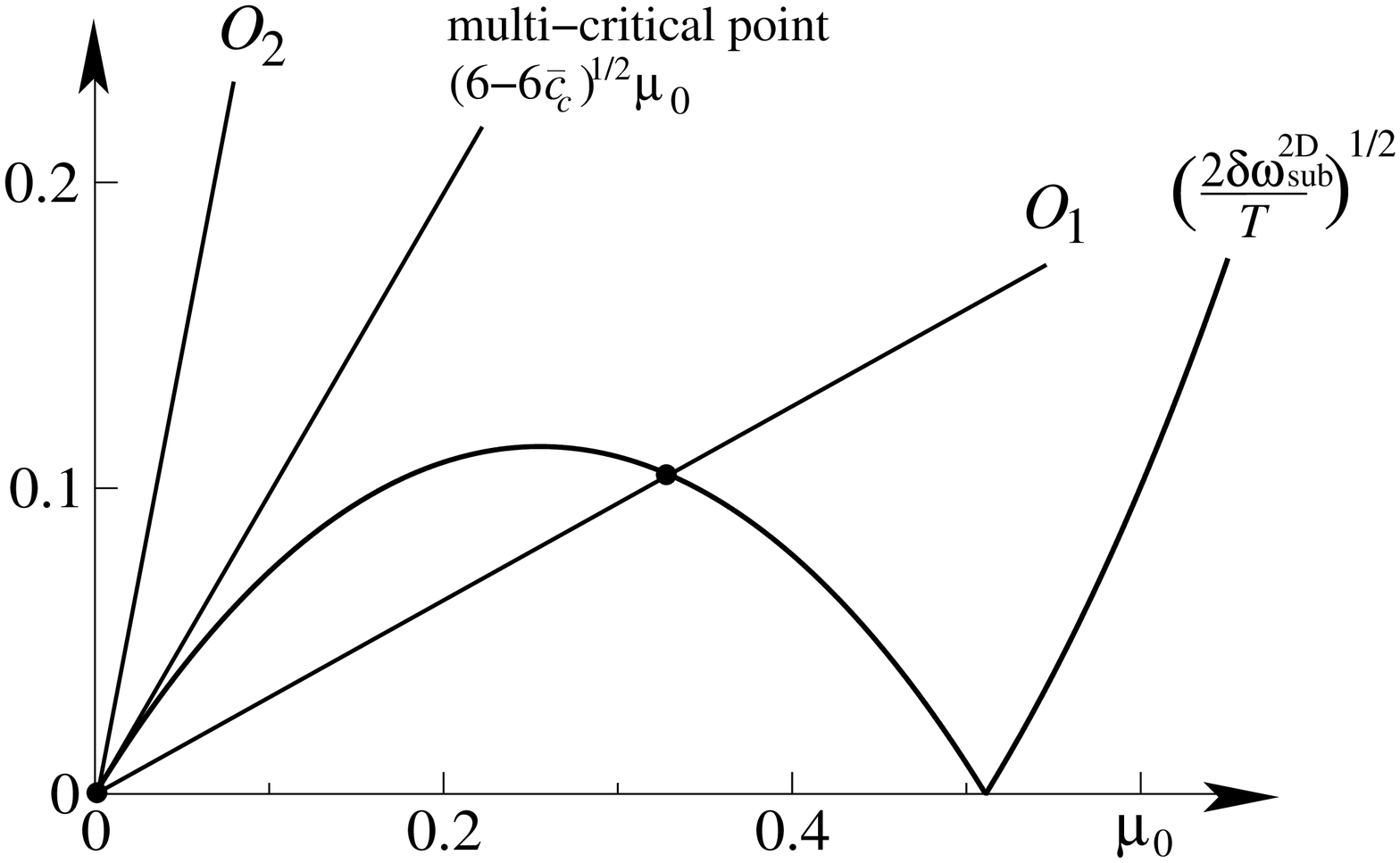}
    \caption[Graphical solution for $\mu_0$]
   {Graphical solution of (\ref{surfcond}) for $\mu_0$. 
   We plot separately the RHS of (\ref{surfcond}), 
   $\sqrt{2\delta\omega^{\rm\scriptscriptstyle 2D}_{\rss sub}(\mu_0)/T}$, 
   and the LHS, $\ell G \mu_0$, for different values of the slope $\ell G$. 
   For $\ell G < \sqrt{6(1-\bar c_c)}$ we find an intersection point at 
   a $\mu_0>0$ (besides the solution $\mu_0=0$). This finite value 
   is the residual order parameter on the surface for the $O_1$ scenario 
   (surface non-melting). For $\ell G = \sqrt{6(1-\bar c_c)}$ the straight 
   line is tangent to 
   $\sqrt{2\delta \omega^{\rm\scriptscriptstyle 2D}_{\rss sub}(\mu_0)/T}$ 
   at $\mu_0=0$, the finite solution has disappeared, and only $\mu_0=0$ 
   remains (multi-critical point). For $\ell G > \sqrt{6(1-\bar c_c)}$ 
   the only solution is at $\mu_0=0$ ($O_2$ scenario, surface melting).
   }
 \label{fig:sqrt}
\end{figure}

\begin{figure}[t]
\centering
    \includegraphics [width= 8 cm] {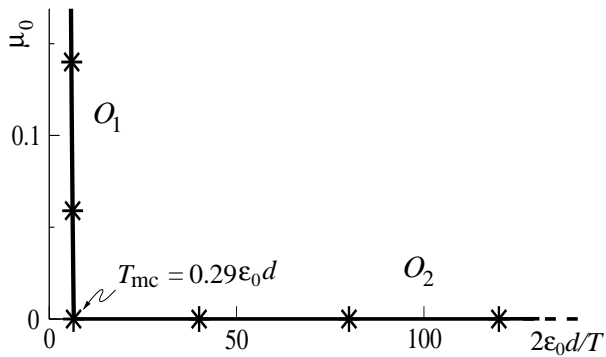}
    \caption[Values of $\mu_0$ at the transition. Comparison
    of analytical and numerical approaches]
   {Residual value of the order parameter at the surface: 
   analytic (solid line) from (\ref{surfcond}) and full numerical solution (stars)
   see Section\ \ref{sec:surfnum}. For $T<T_\mathrm{mc}$ the surface 
   undergoes a continuous transition ($O_2$), whereas for 
   $T>T_\mathrm{mc}$ the order parameter at $z=0$ still exhibits 
   a residual jump at melting.}
 \label{fig:mu0}
\end{figure}

The analysis of the boundary value problem (\ref{eqmotion}) 
with (\ref{bc}) follows the one in Ref.\ \onlinecite{lipowsky83}. 
Combining  the expression 
for the `conserved energy' (originating from the bulk equation 
(\ref{eqmotion}))
\begin{equation}
\ell\, \mu_z^\prime = 
\sqrt{\frac{2 \delta \omega^{\rm\scriptscriptstyle 2D}_{\rss sub}(\mu_z)}{T}}
\end{equation}
with the boundary condition (\ref{bc}), we find an algebraic relation 
which determines the value of the order parameter $\mu_0$ 
at the surface, 
\begin{equation}\label{surfcond}
   \mu_0 \ell G = \sqrt{\frac{2 \delta\omega^{\rm
   \scriptscriptstyle 2D}_{\rss sub}(\mu_0)}{T}}.
\end{equation}
The liquid surface $\mu_0=\mu_\mathrm{l}=0$ is always a solution 
and we deal with a continuous surface melting ($O_2$ scenario) if it is the 
only one. Once a second solution with $\mu_0>0$ is present, 
the surface undergoes a discontinuous ($O_1$) transition, see Fig.\ \ref{fig:sqrt}.
Using the full expression for the potential on the RHS of (\ref{surfcond}),
we find that (\ref{surfcond}) admits two solutions for large $T$ 
(small fields $B$) and only the $\mu_0 =0$ solution for small
temperatures (large magnetic fields), cf.\ the solid line in Fig.\ \ref{fig:mu0}. 
Since both the continuous and the discontinuous melting scenarios are 
present, a multi-critical point separating the two different kinds of transitions 
must exist. The equation which locates the critical point derives 
from (\ref{surfcond}) by considering the quadratic expansion of the 
potential $\delta \omega^{\rm \scriptscriptstyle 2D}_{\rss sub}(\mu)/T
\approx 3(1-\bar{c}_c)\mu^2$ (cf.\  Fig.\ \ref{fig:sqrt})
\begin{align}
  \frac{\ell(T_\mathrm{mc}, B_\mathrm{mc}) 
  G(B_\mathrm{mc})}{\sqrt{6(1-\bar{c}_c)}} 
  & = 1 \nonumber \\ 
  \qquad\longrightarrow (T_\mathrm{mc}, B_\mathrm{mc}) & \approx 
  (0.29\,\varepsilon_0d, 0.007\, B_\lambda).
\label{mc_pos}
\end{align}
For $T<T_\mathrm{mc}$ the surface undergoes a continuous transition
and acts as a nucleus for the liquid phase, preventing the 
appearance of the solid metastable phase. On the other hand, for large 
temperatures $T>T_\mathrm{mc}$ the order parameter at $z=0$ still 
undergoes a residual jump and the double-sided hysteretic behavior 
is restored, see Figs.\ \ref{fig:dftsurf1} and \ref{fig:dftsurf2}.

\subsubsection{Multi-critical point at high-fields}

At high magnetic fields the layers melt independently 
following a first-order type 2D-melting scenario. The order
parameter $\mu_0$ in the topmost layer then undergoes a finite
jump and the surface non-melting ($O_1$) scenario applies.
The presence of a discontinuous $O_1$ regime at high fields 
implies the existence of a second multi-critical point.  
While our analytical approach is not applicable anymore,
since it is based on the gradient expansion which is not justified 
in this  regime, numerically we have found clear indications of a 
finite jump at high fields ($B\approx 10\, B_\lambda$). However, a 
more elaborate version of the DFT is required for an accurate 
determination of this multi-critical point. 
In particular, approaching the melting temperature of each 2D lattice, 
the higher-order peaks in the OCP correlation function $c^{\rss 2D}(K)$ 
become important ($K_n > G$). Hence, higher components in the Fourier 
expansion of the density have to be retained \cite{singh91}, in order to 
obtain a more precise description of the high-field regime. 

\subsection{Analysis of the continuous surface 
melting and of the multi-critical point}

The fact that a discontinuous bulk transition may turn
continuous at the surface is somewhat non-trivial. To
obtain a better insight into the mechanism of the
continuous surface melting, we present here an 
alternative analysis. We describe the surface-melting 
process in the language of the entry of the liquid through 
the boundary. The appearance of a liquid layer 
at the surface, while the bulk remains
solid, implies the existence of an interface between
the two phases, which is described by a soliton-like
profile of the order parameter.  The entry of the
liquid can happen in two different ways: {\it i)}
the soliton can slide smoothly from the boundary at
$T_\mathrm{m}$ (surface melting, $O_2$) or {\it ii)}
it starts entering the system but remains pinned
near the surface at $T_\mathrm{m}$ (surface non-melting,
$O_1$).  To distinguish these two scenarios we need to
calculate the energy of the soliton as a function of its (half-height) 
position $z_\mathrm{s}$ away from the surface and the 
resulting pinning energy. We do this within the local theory 
described by the functional
\begin{equation}
     \frac{\delta \omega}{T}\! =\! \int \frac{dz}{d}   
    \Bigl[\frac{\delta \omega^{\rss 2D}_{\rss sub}(\mu_z)}{T}+
    \frac{\ell^2}{2}\Bigl(\frac{d\mu_z}{dz}\Bigr)^2 
    + \frac{\ell^2 G}{2} \mu_z^2\delta(z)\Bigr].
   \label{local_surf}
\end{equation}
The saddle point equation of (\ref{local_surf}) reproduces
the equation of state (\ref{eqmotion}) with the boundary 
condition (\ref{bc}) at $z=0$ (cf. Eqs.\ (\ref{eq_land}) and (\ref{bc_land})).

\begin{figure}[t]
\centering
    \includegraphics [width= 8 cm] {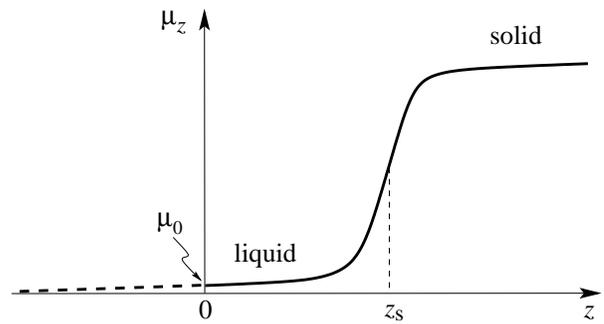}
    \caption[Surface melting: solid-liquid interface]
   {Sketch of a solid-liquid interface placed at a distance 
   $z_\mathrm{s}$ away from the surface.
   The surface destabilizing term $E_\mathrm{surf}(z_\mathrm{s})$ of Eq.\ 
   (\ref{e_surf}) favors the entry of the liquid-solid interface at the surface, 
   producing a repulsive surface potential on the soliton. 
   On the other hand, the energy cost to push the soliton into the system 
   generates an attractive surface potential $E_\mathrm{sl}(z_\mathrm{s})$, 
   cf.\ (\ref{eslz}). Depending on which term is dominant, either a continuous 
   or a discontinuous surface melting transition is realized.  }
 \label{fig:soliton_ws}
\end{figure}

As in the previous section, we concentrate on the case
when the order parameter at the surface is small. 
Since the bulk solution remains valid also in the vicinity 
of the surface, we can study the problem within a variational 
scheme, taking the bulk soliton  $\mu^\mathrm{sl}_z(z_\mathrm{s})$
as a convenient variational function  and displacing
it rigidly at different distances $z_\mathrm{s}$ from
the surface (`sl' stands for solid-liquid interface),
see Fig.\ \ref{fig:soliton_ws}.

We proceed with the estimate of the energy due to a solid-liquid 
interface at $z_\mathrm{s}$, which is given by $\delta\omega
[\mu_z^\mathrm{sl}(z_\mathrm{s})]$.  In the following,
we want to calculate the soliton energy not only
at the bulk transition temperature $T_\mathrm{m}$
but also at temperatures $T$ close by.  However,
for $T\neq T_\mathrm{m}$, the free energy
$\delta\omega[\mu_z^\mathrm{sl}(z_\mathrm{s})]$
is infinite due to the contribution from the
bulk solid $\delta \omega ^{\rss 2D}_{\rss
sub}(\mu_\mathrm{s})\neq 0$.  In order to obtain a
finite energy, which correctly estimates the soliton
energy $E_\mathrm{sl}^\mathrm{surf}(z_\mathrm{s})$
in the presence of the surface we have to subtract
this infinite contribution. We therefore define
\begin{align}
   E_\mathrm{sl}^\mathrm{surf}(z_\mathrm{s})
   & \equiv \delta \omega[\mu_z^\mathrm{sl}(z_\mathrm{s})] - 
  (L/d)\delta \omega_{\rss sub}^{\rss 2D}(\mu_\mathrm{s},T) 
  \label{total_ens}
\end{align}
where $L\to \infty$ is thickness of the sample. 
Inserting the function $\mu_z^\mathrm{sl}(z_\mathrm{s})$ in 
(\ref{local_surf}), the third term is simple to calculate and 
yields the surface energy
\begin{equation}\label{e_surf}
  E_\mathrm{surf}(z_\mathrm{s}) = T\,\frac{\ell^2 G}{d}
  \frac{[\mu_0^\mathrm{sl}(z_\mathrm{s})]^2}{2}.
\end{equation}
The first two terms in (\ref{local_surf}) give the energy of the soliton
in the semi-infinite system. For $T < T_\mathrm{m}$, the presence
of a liquid layer of thickness $\sim z_\mathrm{s}$ produces a
contribution linear in $z_\mathrm{s}$, due to the difference 
between the liquid and solid free energies, 
$-\delta \omega_{\rss sub}^{\rss 2D}(\mu_\mathrm{s}, T)>0$
(we have written the additional argument $T$ in 
$\delta \omega_{\rss sub}^{\rss 2D}$ to make explicit its temperature 
dependence). Subtracting as in (\ref{total_ens}) the infinite contribution 
from the solid phase, this energy becomes 
$-[(z_\mathrm{s}-z^*)/d]\delta \omega_{\rss sub}^{\rss 2D}(\mu_\mathrm{s},T)$, 
with $z^*$ an irrelevant constant. The remaining contribution to 
the soliton energy is due to the interface  $E_\mathrm{sl}(z_\mathrm{s})$, 
i.e., the non-constant part of the order-parameter profile, and consists of both
potential, $\delta\omega^{\rss 2D}_{\rss sub}$, and elastic, $\propto (\mu'_z)^2$, 
energies. Combining these two terms, we find (see Eq.\ (\ref{esl0}))
\begin{align}\label{eslz}
  E_\mathrm{sl}( z_\mathrm{s}) & = \frac{\ell T}{d} 
  \int_{\mu^\mathrm{sl}_0(z_\mathrm{s})}^{\mu_\mathrm{s}}
  d\mu\sqrt{\frac{2\delta\omega^{\rss 2D}_{\rss sub}(\mu)}{T}}\nonumber \\
  & =
  E_\mathrm{sl} - \frac{\ell T}{d} 
  \int_{0}^{\mu^\mathrm{sl}_0(z_\mathrm{s})} 
  d\mu \sqrt{\frac{2\delta\omega^{\rss 2D}_{\rss sub}(\mu)}{T}} \nonumber \\
  & \approx E_\mathrm{sl} - \sqrt{6(1-\bar{c}_c)}\,T\, \frac{\ell}{d}
  \frac{[\mu_0^\mathrm{sl}(z_\mathrm{s})]^2}{2}.
\end{align}
In the last line of (\ref{eslz}) we have expanded the 2D potential for 
small values of $\mu$. Since for a finite $z_\mathrm{s}$ the soliton 
has not fully entered in the system, its energy is less than the total 
soliton energy $E_\mathrm{sl}$ in (\ref{ensol0}) due to the missing tail, 
see Fig.\ \ref{fig:soliton_ws}. One sees that, whereas $E_\mathrm{sl}$ 
penalizes the entrance of the soliton and thus favors the solid phase,  
$E_\mathrm{surf}$ promotes the formation of the liquid phase at the surface.  
Combining all terms, we obtain the soliton energy (\ref{total_ens})
\begin{align}
  E_\mathrm{sl}^\mathrm{surf}( z_\mathrm{s}) &
  \approx E_\mathrm{sl} 
  +\frac{l}{\bar\rho} \frac{T_\mathrm{m}-T}{T_\mathrm{m}}z_\mathrm{s} 
  \nonumber \\
 &\quad + \frac{\ell T_\mathrm{m}}{2d}  
 \Bigl[\ell G - \sqrt{6(1-\bar{c}_c)}\Bigr] 
 [\mu^\mathrm{sl}_0(z_\mathrm{s})]^2,\label{en_sol_s}
\end{align}
where we have expanded the solid free energy term about 
$T_\mathrm{m}$, $\delta \omega^{\rss 2D}_{\rss sub} (\mu_\mathrm{s},T)
 \approx \partial_T  \delta \omega_{\rss sub}^{\rss 2D} 
 (\mu_\mathrm{s},{T_\mathrm{m}}) (T-T_\mathrm{m})$ and we have 
 used the definition of the latent heat $l = T_\mathrm{m} \Delta s = 
 - (\bar\rho/d)T_\mathrm{m} \partial_T \delta \omega_{\rss sub}^{\rss 2D} 
 (\mu_\mathrm{s},T_\mathrm{m})$. 

Below melting, the term $\propto z_\mathrm{s}$ forbids the entry of the 
soliton into the bulk. At the melting transition $T_\mathrm{m}$, the third
term plays the crucial role with a finite order parameter at the surface 
$\mu^\mathrm{sl}_0(z_\mathrm{s})$ leading to a positive or a negative 
energy contribution depending on the sign of the prefactor 
$\propto [\ell G - \sqrt{6(1-\bar{c}_c)}]$. For a positive value 
($\ell G > \sqrt{6(1-\bar{c}_c)}$), the minimal energy is achieved by 
$\mu_0^\mathrm{sl}(z_\mathrm{s})=0$. As a consequence, the soliton 
enters completely inside the sample at  $T_\mathrm{m}$ (surface melting). 
On the other hand, for $\ell G < \sqrt{6(1-\bar{c}_c)}$, 
the entry of the interface costs a positive energy and, thus, the
soliton remains pinned at the surface at $T_\mathrm{m}$. 
The two different behaviors are separated by the multicritical point,
whose location is given by
\begin{equation}
  \ell(T_\mathrm{mc}, B_\mathrm{mc}) G(B_\mathrm{mc}) 
  = \sqrt{6(1-\bar{c}_c)}.
\end{equation}
This equation coincides with (\ref{mc_pos}).

\begin{figure}[t]
\centering
    \includegraphics [width= 7.5 cm] {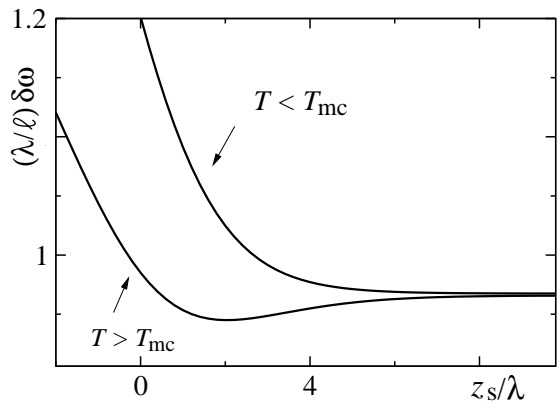}
    \caption[Energy as a function of the soliton position]
   {Energy as a function of the soliton position $z_\mathrm{s}$.
   Note the minimum for
   $T = 0.33\, \varepsilon_0 d > T_\mathrm{mc}$
   ($O_1$, pinned soliton) which has disappeared at
   $T = 0.28\, \varepsilon_0 d < T_\mathrm{mc}$
   ($O_2$, depinned soliton).}
 \label{fig:soliton_en}
\end{figure}

The appearance of the multi-critical point $(T_\mathrm{mc}, 
B_\mathrm{mc})$ along the melting line can be interpreted as a 
surface-depinning transition of the solid-liquid interface (soliton). 
To find the precise pinning location of the soliton in the $O_1$ regime 
a more detailed analysis is required, accounting for the higher order 
terms in (\ref{eslz}).
We have calculated this potential numerically in both the surface melting and 
surface non-melting regimes, using the following scheme: we place 
a bulk soliton at different distances from the surface and evaluate numerically the 
total non-local energy (\ref{free2a}) as a function of $z_\mathrm{s}$. 
In Fig.\ \ref{fig:soliton_en}, we present two different curves,
one at $T = 0.33\, \varepsilon_0 d > T_\mathrm{mc}$ and another 
at $T = 0.28\, \varepsilon_0 d < T_\mathrm{mc}$. 
While for higher temperatures the potential exhibits a stable minimum 
close to the surface where the soliton remains pinned, upon
decreasing the temperature the minimum moves deeper into the bulk 
and disappears altogether at $T_\mathrm{mc}$. 

When the surface melts continuously, we can characterize the transition 
by specific critical exponents, cf.\ Ref.\ \onlinecite{lipowsky83}.
In particular, we can look at how the soliton position $z_\mathrm{s}$ depends 
on  $T$.  Here, we need the function 
$\mu^\mathrm{sl}_0(z_\mathrm{s})$, i.e., the relation between 
the value of the order parameter on the surface and the soliton 
position $z_\mathrm{s}$. For a soliton which is well inside the 
sample, $\mu^\mathrm{sl}_0(z_\mathrm{s})$ can be approximated as
(cf.\ (\ref{eqmotion}))
\begin{equation}\label{mu0z}
  \mu^\mathrm{sl}_0(z_\mathrm{s}) \approx e^{-G_\mathrm{+} 
  \sqrt{r} z_\mathrm{s}}.
\end{equation}
The position of the soliton is found from the minimum of (\ref{en_sol_s})
\begin{align}
  z_\mathrm{s}(T) & \approx -\frac{1}{2G_+\sqrt{r}}
  \ln\left[ \frac{l d (T_\mathrm{m}-T)}{2\sqrt{r}G_+\bar\rho \ell T^2_\mathrm{m}
  (\ell G - \sqrt{6(1-\bar{c}_c)})}\right] \nonumber \\
  & \sim 0.5\ell\, |\ln [(1-T/T_\mathrm{m})/(\ell G - \ell G|_{\rss mc})]|. \label{z_t}
\end{align}
Hence, the soliton slides into the system logarithmically
\cite{Lipowsky1982} with $T \to T_\mathrm{m}$. Next, we study how the residual 
order parameter on the surface goes to zero. Combining 
(\ref{z_t}) together with (\ref{mu0z}), we find that
\begin{equation}
  \mu_0(T) \sim (1-T/T_\mathrm{m})^{1/2}.
\end{equation}
These results are standard in the theory of surface melting when only 
short-range interactions are present \cite{Lipowsky1982}.

For a continuous surface-melting transition the soliton propagates into 
the bulk at $T_\mathrm{m}$, leading, in a semi-infinite system, to the 
coexistence of the liquid and the solid. The interface deep in the bulk 
produces the maximum free energy cost, 
$E_\mathrm{sl}^\mathrm{surf} (z_\mathrm{s} \to \infty) = E_\mathrm{sl}$ 
(see Sec.\ \ref{sec:sol-liq}). This energy is only an interface energy and, hence,
its contribution vanishes in the thermodynamic limit.  
In realistic finite systems, one has to account for the effect of the opposite surface. 
Approaching $T_\mathrm{m}$ from below, both surfaces undergo a continuous 
melting transition. The two surfaces act like two nucleation points for the liquid phase, 
giving rise to two opposite solitons. The system is then composed of a sequence of 
liquid-solid-liquid regions. At $T_\mathrm{m}$ the two solitons merge 
and the intermediate solid domain vanishes altogether. 
Hence, the solid cannot be overheated above the melting temperature.

\section{Numerical analysis}\label{sec:surfnum}

In order to check the accuracy of our analytical approach, we
have carried out a numerical solution of the saddle-point equations 
(\ref{eqmotion_0}).  As a preliminary step, we write (\ref{eqmotion_0}) 
in a slightly different form which is more convenient for our numerical 
study. From (\ref{min1}), we obtain that at the saddle-point the solutions 
for $\xi_z$ and $\mu_z$ are related by
\begin{equation}\label{xiz0}
  \xi_z = \int_0^{\infty} \frac{dz'}{d}\, \bar{c}_{z,z'}\mu_{z'}.
\end{equation}
Combining this expression with the relation (cf.\ (\ref{ximu}))
\begin{equation}\label{muphixi}
  \mu_z = \Phi'(\xi_z)/6,
\end{equation}
where the function $\Phi$ is defined in (\ref{phixi}), we obtain 
a set of integral equations which determine the order-parameter 
profile $\mu_z$
\begin{equation}
  6 \mu_z = 
   \frac{\ds \int_v d^2{\bf R}\, g({\bf R}) 
  \exp\Bigl[ \int_0^{\infty} \frac{dz'}{d}\, 
  \bar{c}_{z,z'}\mu_{z'} g({\bf R})\Bigr]}{\ds \int_v d^2{\bf R}\, 
  \exp\Bigl[\int_0^{\infty} \frac{dz'}{d}\, 
  \bar{c}_{z,z'}\mu_{z'}g({\bf R})\Bigr]},
\label{saddle2}
\end{equation}
where we have written $\Phi'$ explicitly. Our numerical solution 
is based on the recursive solution of (\ref{saddle2}).

We first discretize the $z$ axis in $N=1000$ values $\{z_i\}$ with a
fixed distance $z_i-z_{i-1}=\Delta z = 0.04\,\lambda$ (for $\Gamma > 40$ 
we use a smaller step size $\Delta z = 0.01\,\lambda \approx d$,
since the soliton interface becomes sharper).   
We start from a constant solid phase and initialize the values of $\{\mu_i\}$ 
as $\mu_i=\mu_{z_i}=\mu_\mathrm{s}$, for any $i$.
Then, from Eq.\ (\ref{xiz0}), we derive the molecular field profile $\xi_i$ 
in correspondence to the values $\{z_i\}$. We calculate the RHS of equation 
(\ref{muphixi}) for $i=1,\dots,i_\mathrm{max}$, while keeping the last values 
($i = i_\mathrm{max}+1,\dots,1000$) unchanged, and obtain the new 
$\mu^\mathrm{n}_i = 6 \Phi'(\xi_{z_i})$. 
We compare $\{\mu^\mathrm{n}_{z_i}\}$ with $\{\mu_{z_i}\}$ and
if both the inequalities $\mu^\mathrm{n}_0- \mu_0 < 10^{-5}$ and 
$(1/N)\sum_i (\mu^\mathrm{n}_{z_i}- \mu_{z_i}) < 10^{-5}$ are satisfied, 
we accept $\{\mu^\mathrm{n}_{z_i}\}$ as the order-parameter profile. 
Otherwise, the procedure is iterated recursively until a stable solution 
(fixed point) is reached. We take $i_\mathrm{max}=750$; for this value 
we have checked that the connection between the numerical solution 
for $i \leqslant i_\mathrm{max}$ and the constant bulk value for 
$i>i_\mathrm{max}$ is smooth (we find a small jump at $i_\mathrm{max}$,
$(\mu_{i_\mathrm{max}+1}- \mu_{i_\mathrm{max}})/\mu_{i_\mathrm{max}} 
\approx 10^{-5}$).  Usually convergence is obtained after a reasonable
number of iteration ($< 100$), however in the proximity of a continuous 
surface transition the convergence becomes slow.
This critical slowing down makes it difficult to track the sliding of the soliton inside
the bulk. To avoid this problem, for these critical cases we start our iteration 
from a more convenient initial state. Instead of initiating 
the profile in the bulk homogenous solid, we chose to start from a bulk soliton 
at the position which minimizes the total surface free energy 
(cf.\ Fig.\ \ref{fig:soliton_en} and discussion in the last section). 
In this case the convergence is extremely rapid. 

\begin{figure}[t]
\centering
    \includegraphics [width= 7.1 cm] {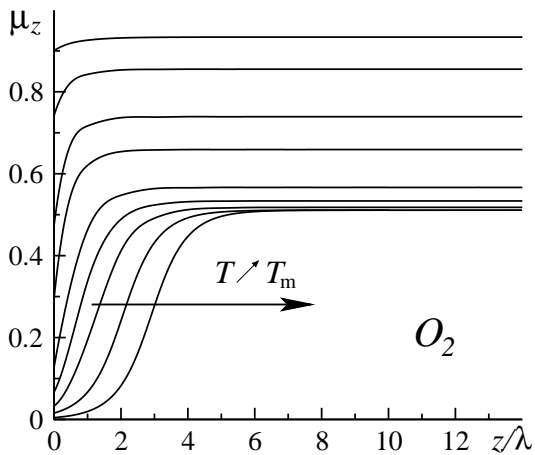}
   \caption[Numerical profiles of the order parameter: surface melting $O_2$
   scenario]
   {Numerical solutions of the order-parameter
   profile $\mu_z$ at the surface, for different 
   values of the external magnetic field at 
   $\Gamma = 2\varepsilon_0 d/ T= 25$, 
   corresponding to $T = 0.08 \varepsilon_0d$.
   At this temperature 
   the bulk melts at  $B_\mathrm{m}/B_\lambda \approx 0.1528$. 
   The lines from top to bottom correspond
   to the magnetic fields with values $B/B_\lambda$ = 0.01,
   0.05, 0.1, 0.125, 0.145, 0.15, $B_\mathrm{m}/B_\lambda- 10^{-4}$, 
   $B_\mathrm{m}/B_\lambda- 10^{-5}$, 
   $B_\mathrm{m}/B_\lambda- 10^{-6}$. 
   While approaching this value the soliton slides into the bulk
   (see the bottom three lines). 
   The surface undergoes an $O_2$ type transition, i.e., 
   $\mu_0$ approaches zero (liquid phase) continuously.  }
 \label{fig:dftsurf1}
\end{figure}

\begin{figure}[t]
\centering
    \includegraphics [width= 7.3 cm] {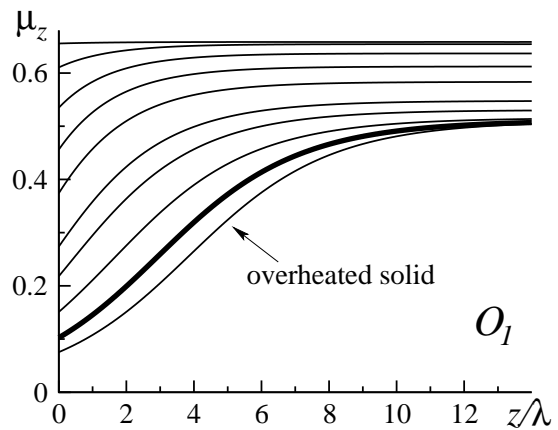}
    \caption[Numerical profiles of the order parameter: 
    surface non-melting $O_1$ scenario]
   {
   Numerical solutions of the order-parameter profile $\mu_z$ near the surface 
   at $T = 0.33\, \varepsilon_0d$ ($\Gamma = 6$) for different values of the 
   magnetic field. The freezing field is $B_\mathrm{m}/B_\lambda 
   \approx 0.002396$ (thicker line). The lines from top to bottom correspond
   to the magnetic fields with values $B/B_\lambda$ = 0, 0.0001, 0.0005, 0.001, 
   0.0015, 0.002, 0.0022, 0.00235, $B_\mathrm{m}/B_\lambda$, 
   $B_\mathrm{m}/B_\lambda + 10^{-5}$. The order parameter at 
   the surface jumps discontinuously to zero; hence, the surface transition 
   is in the class $O_1$. In this case, the surface does 
   not preclude the appearance of the overheated solid.
   Numerically it is possible to obtain a non uniform solution 
   even at magnetic fields $B > B_\mathrm{m}$ 
   larger than the freezing one (cf.\ lowest thin line), 
   corresponding to a metastable configuration (overheated solid).
   }
 \label{fig:dftsurf2}
\end{figure}

In Figs.\ \ref{fig:dftsurf1} and  \ref{fig:dftsurf2}, we show two different 
examples of the order-parameter profile for two different values of the 
temperature:  {\it i)} $T= 0.08\ \varepsilon_0 d <T_\mathrm{mc}$  in 
Fig.\ \ref{fig:dftsurf1}, where the surface undergoes a continuous 
transition and {\it ii)} $T= 0.33\ \varepsilon_0 d  > T_\mathrm{mc}$ in 
Fig.\ \ref{fig:dftsurf2} where a discontinuous surface transition takes place.
For the $O_2$ transition at $T= 0.08\,\varepsilon_0 d$, 
we show the order-parameter profiles for different values of the 
magnetic field $B$, increasing from top to bottom.  At low magnetic 
fields, the profile is almost constant (cf.\ the topmost line), 
since the surface kernel $\delta \omega^\mathrm{s}$ is negligible and 
the system is translationally invariant. 
Going to larger values of the magnetic field, the 
modifications of the interlayer potential on the surface 
become relevant, reducing the value of the order parameter at 
the surface. Upon increasing the magnetic field further towards the
thermodynamic melting transition, the vortex density
modulation becomes vanishingly small close to the
surface.  The numerical solution shows that the order
parameter at the surface goes to zero (liquid) continuously.  
Hence, the surface assists the penetration of the liquid phase by 
the formation of a quasi-liquid nucleus.  This continuous transition 
on the surface eliminates the hysteresis above the melting transition, 
by preventing the appearance of a metastable overheated solid phase.


At larger temperatures  $T= 0.33\, \varepsilon_0 d >T_\mathrm{mc}$ 
(Fig.\ \ref{fig:dftsurf2}), 
the surface undergoes a discontinuous transition, although with a 
reduced jump in comparison with the bulk. 
Again starting from low magnetic fields (topmost line) the order
parameter is constant, due to the smallness of the surface 
destabilizing potential. Increasing $B$, the value of $\mu_0$ decreases,
still showing a larger suppression than in the bulk. 
However, the surface potential is not strong enough to push $\mu_0$
to zero and at melting (thick line) the surface still exhibits a finite 
order parameter. The transition to the homogeneous liquid phase 
then occurs via a finite global jump including the surface value $\mu_0$ 
as well. This discontinuous transition is compatible with the appearance 
of the metastable phase. Indeed, numerically it is possible to obtain a 
non-uniform solution even at magnetic fields which are
larger than the freezing field (lowest line in Fig.\ \ref{fig:dftsurf2}). 

Finally, we check the accuracy of our analytical approach in estimating the 
location of the multi-critical point $(T_\mathrm{mc}, B_\mathrm{mc})$. 
We plot the residual value of the surface order 
parameter at melting as a function of temperature in Fig.\ \ref{fig:mu0}, 
together with the solution of (\ref{surfcond}). 
For a continuous surface melting transition the order parameter is exactly zero at
melting. Numerically, we associate
the value $\mu_0=0$ to situations where the soliton potential does not show a 
stable minimum as a function of $z_\mathrm{s}$ (see Fig.\ \ref{fig:soliton_en}).
Otherwise, we estimate the finite value of $\mu_0$ within the iterative solution of the
saddle point equations, which we have described above. The agreement of 
the numerical and analytical results is excellent, in particular in the vicinity 
of the multi-critical point.

\section{Conclusions}

In this paper, we have analyzed the impact of an
$ab$-surface on the melting transition of the pancake
vortex lattice.  We have adapted the DFT-substrate
approach\cite{De-Col2006} to include the presence of the surface. 
We have found that for intermediate values of the
magnetic field, the surface undergoes a continuous
melting transition and assists a smooth propagation
of the liquid into the bulk. This result reveals
the origin of the asymmetric hysteresis at the
melting transition observed by Soibel {\it et al.}
\cite{soibel00} as the consequence of free surfaces
which can act as nucleation sites for the liquid phase.
Moreover, we have found that at low and high magnetic
fields the surface transition turns discontinuous
as in the bulk. The continuous and 
discontinuous transitions at the surface are separated
by a multi-critical point. While the precise location of
the high-field multi-critical point goes beyond the limits of 
validity of our analysis, we have determined the position of
the low-field multi-critical point by means of an analytical solution 
of the DFT equations and numerically confirmed the result of 
our analysis.

An effect which we have not included in our study is the impact of the 
reentrance of the melting line at low magnetic fields. 
At low magnetic fields 
the interaction between full {\it vortex lines} is strongly
screened. This leads to a softening of the vortex lattice with exponentially 
small shear and compression moduli  and to the reentrance of
the melting line towards small temperatures. The effect of a surface 
in this reentrant low-field regime is an interesting problem as well. 
The stray magnetic fields produce an 
algebraic interaction between the tips of the vortex lines\cite{pearl2} instead of 
the exponentially small bulk interaction. This may lead to a highly 
unconventional scenario (proposed in Ref.\ \onlinecite{huse92}) where  
the tips of the vortex lines on the surface are arranged in a regular lattice while 
the vortex system is already melted in the bulk. 
Hence, in this case the surface plays a role which is opposite to the
one played in the standard surface melting scenario, since it favors the appearance 
of the solid instead of the liquid. The impact of this `surface solidification' on our
results remains an interesting open question.

We acknowledge support from the Swiss National Foundation
through MaNEP (ADC) and from the DST (India) through
a Swarnajayanti Fellowship (GIM).

\appendix


\section{Gradient expansion}\label{app:ge}

\subsection{Kernel in Fourier space}

\begin{figure}[t]
\centering
\includegraphics [width=7.3 cm] {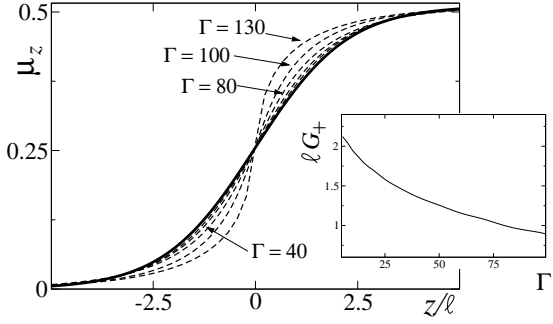}
\caption[Solitons: comparison of local and non-local theories]
{Comparison of the soliton derived within the local approximation 
(thicker line) with the results of the full non-local theory for 
different values of $\Gamma = 6,40,80,100,130$. For $\Gamma = 6$,
upon rescaling $z$ in units of $\ell$, the soliton collapses 
on the solution of the local approximation. 
Little deviations are present at $\Gamma = 40$, where
the soliton is still well approximated by the gradient expansion. 
At higher $\Gamma$'s the difference is appreciable: 
the full non-local theory leads to a shaper interface.
Inset: plot of the ratio between $\ell$ and kernel extension $1/G_+$ as
a function of $\Gamma = 2\varepsilon_0 d/T$. Assuming that 
the gradient expansion is valid for $\ell_\mathrm{e} G_+ 
\approx 4\, \ell G_+\gtrsim 5$, we obtain the limiting value 
$\Gamma \approx 50$, corresponding to $T\approx 0.04\, \varepsilon_0 d$.
}
\label{fig:collapsedsoliton}
\end{figure}

In this Appendix we test the validity of the gradient expansion 
(\ref{freebulkgradient}) for the DFT free energy in the low-field
regime. We start from the expression of the bulk free-energy 
(\ref{free2abulk}) and calculate the saddle point equation
\begin{equation}
   - 6\bar{\alpha} \int \frac{dz'}{d}   
  \bar{f}_{z-z'}
    (\mu_z-\mu_{z'}) =  
    \partial_{\mu_z} \delta\omega^{\rss 2D}_{\rss sub}(\mu_z)/T ,
    \label{eqmotionnl}
\end{equation}
which has to be compared with the corresponding 
equation (\ref{eqmotion0}) originating from the approximated 
local theory. The difference between the two left-hand side terms of 
(\ref{eqmotionnl}) and (\ref{eqmotion0}) is more conveniently 
analyzed in Fourier space.
In the case of the local theory, the second derivative term is diagonal 
in Fourier space and its components read
\begin{equation}
  \ell^2 k^2 \mu_k
  .\label{fl}
\end{equation}
In the full non-local theory, the left hand side of 
(\ref{eqmotionnl}) is also diagonal in Fourier space
\begin{equation}
    6\bar\alpha(\bar{f}_k
     - \bar{f}_{k=0}) \mu_k 
    = \frac{\ell^2 k^2}{1+(k/G_+)^2} \mu_k,\label{fnl}
\end{equation}
where we have used $\ell = 12 \bar\alpha/dG_+^3$, cf.\ (\ref{ell0}).
From the comparison of (\ref{fnl}) with (\ref{fl}) it becomes clear that 
for small wave numbers $k \ll G_+ $, i.e., for variations 
on scales larger than $1/G_+$, the gradient expansion 
approximates well the full non local theory. 
A convenient test is to compare the typical length of 
the soliton $\ell_\mathrm{e}\approx 4\,\ell$ (cf. (\ref{elle1}))
with the size of the kernel $1/G_+$. If we choose $\ell_\mathrm{e} 
G_+\approx 5$ as the limiting value for the validity of the expansion, 
we obtain that for $\Gamma \lesssim 50$ (corresponding to 
$T\gtrsim 0.04\, \varepsilon_0 d$ or $B\lesssim 0.5\,B_\lambda$) 
the gradient approximation is justified, 
see the inset of Fig.\ \ref{fig:collapsedsoliton}.
Expanding (\ref{fnl}) to second order, we find that the maximum 
error  is only few percent,  $(k/G_+)^2 \approx 1/(\ell_\mathrm{e} G_+)^2 
\lesssim 0.04$. To confirm this result we have solved numerically 
the integral equation (\ref{eqmotionnl})
imposing the boundary conditions $\mu_{z\to -\infty} = 0$ and 
$\mu_{z\to \infty} = \mu_\mathrm{s}$. The results for different 
values of $\Gamma$ are plotted in Fig.\ \ref{fig:collapsedsoliton}, 
together with the soliton of the local theory (thick line). We have 
rescaled the $z$ axis in units of the elastic length $\ell$.
For $\Gamma < 80$ we obtain a good collapse of the data. 
On the other hand, for large $\Gamma$ the kink in non-local 
theory is sharper than the soliton in the local theory and, therefore, 
the gradient expansion approximates poorly the exact result.

\subsection{Linearised saddle-point equation}

For small values of $\mu_z$, we can expand the potential in the RHS of 
the equation (\ref{eqmotionnl}). Hence, we write, cf.\ (\ref{linf2d}), 
\begin{equation}\label{flin2d1}
  \partial_{\mu_z} \delta\omega^{\rss 2D}_{\rss sub}(\mu_z)/T 
  \approx 6(1-\bar{c}_c) \mu_z,
\end{equation}
where we consider the system at melting, i.e., 
$c^{\rss 2D}_{\rss sub} = \bar{c}_c$ to connect with the discussion 
in Sec.\ \ref{sec:dftsurf}. Inserting (\ref{flin2d1}) and (\ref{fnl}) in 
(\ref{eqmotionnl}), we obtain the equation of `motion' in Fourier space
\begin{equation}
  -\ell^2 k^2\Bigl[1+ \frac{6(1-\bar{c}_c)}{(\ell G_+)^2} \Bigr]\mu_k = 
  6(1-\bar{c}_c)\mu_k.
\end{equation}
We can transform this equation back to real space
\begin{equation}
  (1+ r)\ell^2 \mu_z^{\prime\prime} = 6(1-\bar{c}_c)\mu_z,
\end{equation}
where we have defined the parameter 
$r=d G_{\scriptscriptstyle +}(1-\bar{c}_c)/2\bar\alpha = 
6(1-\bar{c}_c)/(G_+\ell)^2$, as in Eq.\ (\ref{eq_nlt}).
The higher derivatives which are neglected in the gradient expansion
produce a small renormalization of the elastic term $(1+r)\approx 1$.


\begin{thebibliography}{31}
\bibitem{note2}
This term is absent in the gradient 
expansion of the bulk equation, the difference being due to the 
higher-order derivatives $\partial_z^n\mu_z$ that are 
neglected in the gradient expansion (see Appendix \ref{app:ge}).
\end{thebibliography}
\end{document}